%% file: Mullaney.tex
\documentclass[useAMS,usenatbib]{./mnras}
\usepackage{times,mathptm}
\usepackage{lscape}
\usepackage{natbib}
\usepackage{graphicx}
\usepackage{floatflt}
\usepackage{amssymb}

\newcommand{\kms}{km~s$^{-1}$}

\newcommand{\ergs}{ergs~s$^{-1}$}

\newcommand{\whz}{W~Hz$^{-1}$}

\newcommand{\chisq}[1]{$\chi^2$}

\newcommand{\fig}[1]{fig. \ref{#1}}

\newcommand{\Tab}[1]{Table \ref{#1}}
\newcommand{\Fig}[1]{Fig. \ref{#1}}

\newcommand{\hii}{H{\sc ii}}

\newcommand{\comment}[1]{}

\newcommand{\fwhma}{${\rm FWHM_{Avg}}$}

\newcommand{\oiii}{$[$O~{\sc iii}$]\lambda$5007}
\newcommand{\ha}{H$\alpha$}
\newcommand{\hb}{H$\beta$}
\newcommand{\nii}{$[$N~{\sc ii}$]\lambda$6584}
\newcommand{\niis}{$[$N~{\sc ii}$]\lambda$6548,6584}
\newcommand{\oiiis}{$[$O~{\sc iii}$]\lambda$4959,5007}

\newcommand{\fha}{${\rm FWHM_{H\alpha}}$}

\newcommand{\loiii}{${L_{\rm [O~III]}}$}
\newcommand{\lrad}{${L_{\rm 1.4~GHz}}$}

\newcommand{\lagn}{${L_{\rm AGN}}$}

\newcommand{\redd}{${\lambda_{\rm Edd}}$}
\newcommand{\rrad}{${R_{\rm Rad}}$}

\title[AGN gas kinematics]
{Narrow-line region gas kinematics of 24,264 optically-selected AGN: \\
  the radio connection\thanks{Catalogue of multi-component fit data
    available at: http://sites.google.com/site/sdssalpaka}}
\author[J. R. Mullaney et al.]  {J. R. Mullaney$^{1,2}$\thanks{E-mail:
    j.r.mullaney@dur.ac.uk}, D. M. Alexander$^{1}$, S. Fine$^{1}$,
  A. D. Goulding$^{3}$, C. M.  Harrison$^{1}$, \newauthor
  R. C Hickox$^{4}$\\
  $^{1}$Department of Physics, Durham University, South
  Road, Durham, DH1 3LE, U.K.\\
  $^{2}$Laboratoire AIM, CEA/DSM-CNRS-Universit\'{e} Paris Diderot,
  Irfu/Service d’Astrophysique, CEA-Saclay, Orme des Merisiers,\\
  \ \ 91191 Gif-sur-Yvette Cedex, France\\
  $^{3}$Harvard-Smithsonian Center for Astrophysics, 60 Garden Street,
  Cambridge, MA 02138, U.S.\\
  $^{4}$Department of Physics and Astronomy, Dartmouth College, 6127
  Wilder Laboratory, Hanover, NH 03755, USA}
\begin{document}

\date{Date Accepted}


\maketitle

\label{firstpage}

\begin{abstract}
  Using a sample of 24264 optically selected AGNs from the SDSS DR7
  database, we characterise how the profile of the \oiii\ emission
  line relates to bolometric luminosity (\lagn), Eddington ratio,
  radio loudness, radio luminosity (\lrad) and optical class (i.e.,
  broad/narrow line Seyfert 1, Type 2) to determine what drives the
  kinematics of this kpc-scale line emitting gas. Firstly, we use
  spectral stacking to characterise how the average \oiii\ profile
  changes as function of these five variables. After accounting for
  the known correlation between \lagn\ and \lrad, we report that
  \lrad\ has the strongest influence on the \oiii\ profile, with AGNs
  of moderate radio luminosity (\lrad$=10^{23}-10^{25}$~\whz) having
  the broadest \oiii\ profiles. Conversely, we find only a modest
  change in the \oiii\ profile with increasing radio loudness and find
  no significant difference between the \oiii\ profiles of broad and
  narrow-line Seyfert 1s. Similarly, only the very highest Eddington
  ratio AGNs (i.e., $>0.3$) show any signs of having broadened \oiii\
  profiles, although the small numbers of such extreme AGNs in our
  sample mean we cannot rule out that other processes (e.g., radio
  jets) are responsible for this broadening.  The \oiii\ profiles of
  Type 1 and Type 2 AGNs show the same trends in terms of line width,
  but Type 1 AGNs display a much stronger ``blue-wing'', which we
  interpret as evidence of outflowing ionised gas. We perform
  multi-component fitting to the \hb, $[$O~{\sc
    iii}$]\lambda\lambda4959,5007$, $[$N~{\sc
    ii}$]\lambda\lambda6548,6584$ and \ha\ lines for all the AGNs in
  our sample to calculate the proportions of AGNs with broad \oiii\
  profiles. The individual fits confirm the results from our stacked
  spectra; AGNs with \lrad$>10^{23}$~\whz\ are roughly 5 times more
  likely to have extremely broad \oiii\ lines (\fwhma$>1000$~\kms)
  compared to lower \lrad\ AGNs and the width of the \oiii\ line peaks
  in moderate radio luminosity AGNs (\lrad$\sim10^{24}$~\whz). Our
  results are consistent with the most disturbed gas kinematics being
  induced by compact radio cores (rather than powerful radio jets),
  although broadened \oiii\ lines are also present, but much rarer, in
  low \lrad\ systems. Our catalogue of multi-component fits is freely
  available as an online resource for statistical studies of the
  kinematics and luminosities of the narrow and broad line AGN regions
  and the identification of potential targets for follow-up
  observations.
  
\end{abstract}

\begin{keywords}
  Keywords
\end{keywords}

\section{Introduction}
\label{Introduction}

The most successful models of galaxy evolution invoke interactions
between accreting supermassive black holes (hereafter, SMBHs) and
their host galaxies to reproduce the most fundamental properties of
today's galaxy population (e.g., \citealt{DiMatteo05, Bower06}). Many
of these models assume that a fraction of the energy released by an
accreting SMBH (i.e., an active galactic nucleus; hereafter, AGN)
drives gas and dust from the central regions of a galaxy, quenching
star-formation (e.g., \citealt{DiMatteo05, Springel05,
  Hopkins06}).\footnote{This ``quasar-mode'' feedback is distinct from
  the ``radio-mode'', in which the intracluster medium is prevented
  from cooling onto galaxies (where it can form stars) by powerful,
  extended, radio jets produced by a low accretion rate, radio-loud
  AGN.}  However, it is not yet clear what process drives these
galaxy-scale ``outflows'', with both radio jets (either compact or
extended, e.g., \citealt{Nesvadba06, Nesvadba08, Rosario10}) and
radiation pressure (e.g. \citealt{Alexander10}) having been suggested
as viable mechanisms.

Most observational evidence for galaxy-scale outflows has so-far been
based on pointed observations of a handful of objects (e.g.,
\citealt{Nesvadba06, Nesvadba08, Alexander10, Rosario10, Greene11,
  Harrison12}).  These studies, while providing considerable insight
into the scale and detailed kinematics of the outflows, are subject to
strong selection biases (e.g., pre-selected for radio
loudness/luminosity, bolometric luminosity, unique emission line
properties).  Therefore, it has so-far been difficult to place these
outflows in the context of the general AGN and galaxy populations and
establish which AGN properties are most conducive to their production
(i.e., radio jets/cores, high bolometric luminosity, high Eddington
ratio etc.).  In this study we aim to address this by taking the
complementary approach of exploring in coarser detail the gas
kinematics of much larger numbers of AGNs with physical properties
spanning large sections of parameter space.

The primary diagnostics that the above studies used to initially
identify kpc-scale outflows were broad, often blueshifted \oiii\
emission lines in integrated (i.e., 1D) rest-frame optical spectra.
Since \oiii\ is produced through a forbidden transition it is only
emitted by low-density gas.  As such, any broadening or shifting of
the \oiii\ emission lines are the result of strong velocity gradients
or bulk motions (i.e., disturbed kinematics) in the so-called
narrow-line region (hereafter, NLR) of AGNs which can be extended over
kpc-scales (e.g. \citealt{Pogge89}), as opposed to the dense, sub-pc
sized broad-line region.

While broadened (beyond that expected from galaxy rotation) and
asymmetric \oiii\ profiles in AGN spectra have been reported for over
three decades (early studies include, e.g., \citealt{Heckman81,
  Feldman82, Heckman84, Whittle85a}), there remains some uncertainty
as to what physical process cause the gas kinematics that result in
these profiles.  Early studies of small to moderately sized samples of
nearby AGNs (numbering $<150$) reported positive correlations between
\oiii\ profile widths and both \oiii\ and 1.4~GHz radio luminosities
for flat-spectrum radio sources with \lrad$<3\times10^{24}$~\whz
(e.g., \citealt{Heckman84}, also \citealt{Wilson80, Whittle85b}).
However, since \loiii\ is broadly correlated with \lrad (e.g.,
\citealt{Baum89, Rawlings89, Zirbel95, Tadhunter98, Wills04,
  deVries07}), these early studies were unable determine which of
these parameters were more fundamentally linked to the kinematics of
the narrow-line region.  With the availability of large spectroscopic
data sets (e.g., the SDSS and 2dF surveys), more recent studies have
explored how the \oiii\ profile changes across wide areas of parameter
space.  For example, using a sample of 1749 SDSS-selected AGNs,
\cite{Greene05} found that the narrow ``core'' of the \oiii\ line
traced the kinematics of the host galaxy, with any departures
correlating with the Eddington ratio of the AGN (also
\citealt{Bian06}; although see also \citealt{Zhang11}).  However,
while some recent studies have explored, in detail, the links between
gas kinematics and the radio properties for small samples of AGNs
(e.g., \citealt{Wu09, Husemann13}), analysis of much larger samples
are needed to break the degeneracy between radio and \oiii\
luminosities to determine how gas kinematics relate to the radio
properties for the AGN population in general.

In this study, we use the SDSS spectral database to derive the average
\oiii\ line profiles of a large sample (i.e., numbering 24,267) of
optically-selected AGNs at $z<0.4$ and explore how these average
profiles change as a function of key AGN parameters (specifically,
\oiii\ luminosity, Eddington ratio, radio loudness, radio luminosity
and spectral classification $[$i.e., optical Type 1, Type 2$]$).
While the single aperture spectra provided by the SDSS provide no
information regarding the extent of any kinematically disturbed gas,
the SDSS spectral archive provides the statistics needed to determine
how gas kinematics relate to other AGN properties (including the radio
properties) and establish the prevalence of high gas velocities among
different AGN sub-populations.  Furthermore, it is important to stress
that the redshift range of our sample does not span the epoch of peak
star-formation and AGN activity (i.e., $z\sim1-3$;
\citealt{HopkinsBeacom06, Merloni04, Aird10}) when quasar-mode AGN
driven outflows are thought to have had their greatest impact on
galaxy growth.  However, the high signal-to-noise ratios and large
statistics allowed by surveys of the local Universe can give us a
better insight into which physical processes are most capable of
disturbing the gas reservoirs around an AGN.

In the following section we outline our initial source selection
procedure, describe our multi-component line fitting routine and
outline how the results from this routine were used to reclassify and
analyse the AGNs in our sample.  In \S\ref{Results} we describe the
results derived from the average \oiii\ profiles, paying particular
attention to how the \oiii\ profile changes as a function of the
parameters outlined above.  In \S\ref{Individual}, we present the
results from our multi-component fitting routine and calculate the
fractions of local, optically-selected AGNs that show broad \oiii\
lines.  We discuss the implications of our main results in
\S\ref{Discussion} and summarise our findings in \S\ref{Summary}.  We
adopt $H_{0}=71$~km~s$^{-1}$~Mpc$^{-1}$, $\Omega_{\rm M}=0.27$, and
$\Omega_{\Lambda}=0.73$ throughout.

\begin{figure*}
  \begin{center}
    \includegraphics[width=15.0cm]{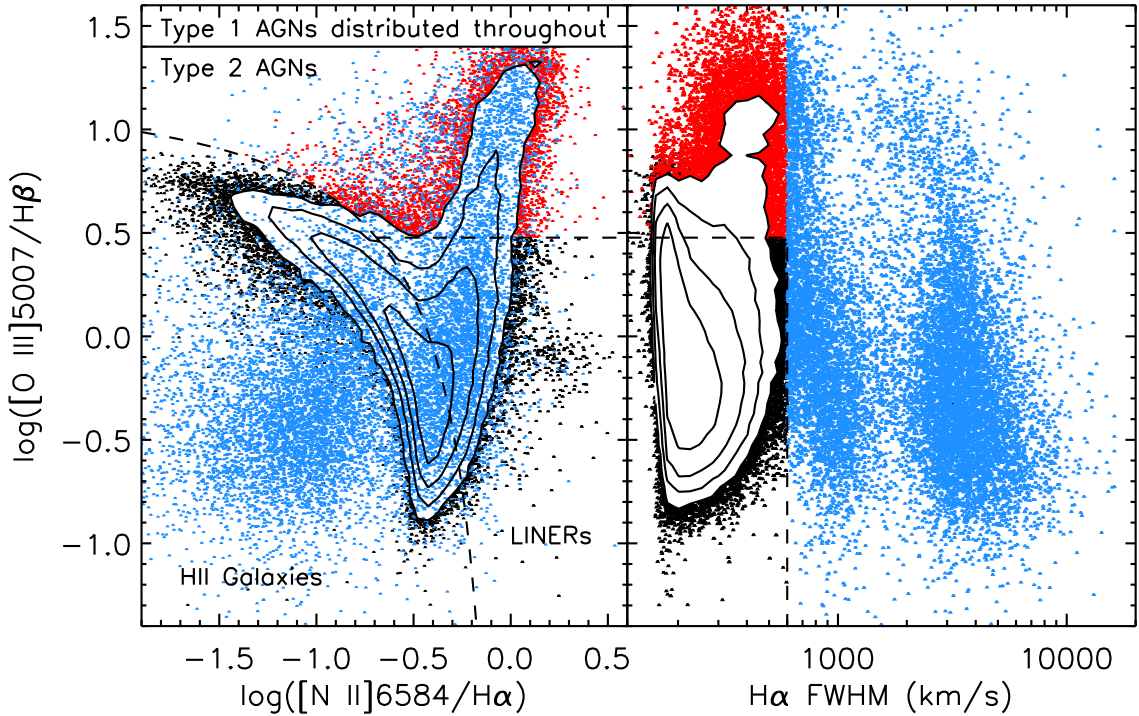}
  \end{center}
  \caption{Plots highlighting the line ratio and \fha\ cuts used to
    identify candidate Type 1 and Type 2 AGNs (shown as blue and red
    points, respectively; black points are used to indicate LINERS and
    H{\sc ii} galaxies) in the SDSS parent sample of emission line
    galaxies.  All values used to create both these plots were taken
    directly from the SDSS database.  {\it Left}: \oiii/\hb\
    vs. \nii/\ha\ BPT diagram of all galaxies in the parent sample.
    Dashed lines indicate the criteria used to discriminate between
    candidate narrow-line objects (i.e., Type 2 AGNs, LINERS and HII
    galaxies; discrimination lines taken from
    \citealt{Ho97,Kauffmann03}). {\it Right}:\oiii/\hb\ vs. \fha\ of
    the same sources.  The width of the \ha\ line was used to
    discriminate between narrow line galaxies (i.e., candidate Type 2
    AGNs, LINERS and H{\sc ii} galaxies) and candidate ``broad line''
    (Type 1) AGNs.  A conservative \fha\ $> 600$~\kms\ cut was used to
    identify candidate Type 1 AGNs.  We note that the bimodal
    distribution of \fha\ (a result of an apparent dearth of AGNs
    around \ha$=2000$~\kms) is a feature produced by the SDSS spectral
    fitting routine.  This bimodality disappears when the \ha\
    profiles are fit with multiple components (see
    \S\ref{FittingRoutine}).  Contours are used in both plots to show
    regions of very high densities of narrow line galaxies (i.e., Type
    2 AGNs, LINERS and H{\sc ii} galaxies).}
  \label{BPT}
\end{figure*}

\section{Sample Selection, Classification and Analysis}
The large number of flux-calibrated galaxy and AGN spectra contained
within the SDSS make it an ideal resource for deriving the average
\oiii\ profiles of optically selected AGNs and quantifying the
proportions of the AGN sub-population that display broad \oiii\
profiles.  However, careful source selection is necessary to remove
contaminants such as \hii\ galaxies from our sample of SDSS AGNs.  We
start this section with a description of how we used the results from
the SDSS spectral fitting routine to efficiently remove those sources
whose optical emission lines are least likely to be the result of
AGN-dominated ionisation processes (e.g. \hii, LINERS etc.).  We then
describe how we used the results of our own emission line fitting
routine to: (a) further refine our AGN sample, removing any remaining
contaminants in the process and (b) measure any broad emission line
components likely missed by the single-component fits carried out by
the SDSS emission line fitting routine.  Next, we explain how we used
the FIRST and NVSS radio coverage of the SDSS field to identify the
radio loud/luminous AGNs in our SDSS sample, allowing us to explore
the potential role of radio AGNs in driving outflows.  Finally, we
describe our spectral stacking procedure which we use as our principal
form of analyses

\subsection{SDSS Query}
\label{Data:Query}
As the main focus of this study is to determine how the profile of the
\oiii\ emission line relates to other key AGN properties, we selected
all extragalactic spectroscopic sources in the SDSS-DR7 catalogue
(\citealt{Abazajian09}) with \oiii\ detected in emission (at
$>3\sigma$) by the SDSS spectral fitting routine.  Since we rely on
emission line diagnostics to identify AGNs (see
\S\ref{Data:Selection}) we required that the \ha, \hb\ and \nii\ lines
are also detected in emission (again, at $>3\sigma$ confidence).  In
ensuring that the \nii\ line is detected, sources at redshifts $z >
0.4$ (at which point \nii\ is shifted out of the wavelength coverage
of the SDSS spectra) are, by default, excluded from the sample.  This
query returned 239,083 SDSS targets.

\subsection{Selection of AGNs}
\label{Data:Selection}
The vast majority of extragalactic emission line sources in the SDSS
database are dominated by star formation rather than AGN activity and
need to be removed from the sample prior to further analysis.  We used
a combination of emission line flux ratios (``BPT'' diagnostics;
\citealt{Baldwin81}) and line widths (as measured by the SDSS spectral
fitting routine) to remove the $\sim$90 per cent of the initial sample
likely dominated by star formation.  We classified all narrow line
objects (defined here as having \fha$<600$~\kms) as being probable
Type 2 AGNs if their \oiii/\hb\ and \nii/\ha\ flux ratios satisfy the
diagnostics outlined in \citeauthor{Kauffmann03}
(\citeyear{Kauffmann03}; see our \Fig{BPT}) and identify all sources
with \fha$>600$~\kms as {\it potential} Type 1 AGNs.\footnote{We rely
  on the FWHM of the \ha\ line, rather than \hb, to ensure we include
  intermediate-class AGNs (e.g. Type 1.8, 1.9 etc.) in our Type 1
  classification} Our choice of \fha\ cut is a compromise between
obtaining as complete a sample of Type 1 AGNs as possible and the
exclusion of H{\sc ii} galaxies and LINERS, whose numbers increase
significantly at \fha$<600$~\kms.  Tests of a 10 per-cent sampling of
these H{\sc ii} and LINERs using our multi-component fitting routine
(see \S\ref{FittingRoutine} and Appendix A) suggest a $\approx0.3$ per
cent Type-1 AGN contamination amongst these non-AGN. As such, we
estimate we are missing $\sim$700 Type 1 AGNs by excluding H{\sc ii}
and LINERs from further analysis.  These ``unidentified'' Type 1 AGNs
will be the focus of a later study.

The above cuts identified 25,670 potential Type 1 and Type 2 AGNs that
were passed to our multi-component fitting routine for further
analysis and possible re-classification.  The remaining 213,413
galaxies are excluded from any further analysis.

\subsection{Multi-Component Fitting Routine}
\label{FittingRoutine}
In the previous subsection we described how we used the results from
the SDSS emission line fitting routine to crudely select samples of
Type 1 and Type 2 AGNs in the SDSS database.  However, the single
gaussian (i.e., one per emission line) fits performed by the SDSS
pipeline cannot account for the complex emission line profiles that
are common among AGN spectra.  Instead, we developed our own
multi-component fitting routine to accurately measure these line
profiles.  We used this routine to re-measure the profiles of the \ha,
\hb, \niis\ and \oiii\ emission lines in all 25,670 spectra identified
in \S\ref{Data:Selection}.  The results from this multi-component
fitting routine were used to (a) obtain more robust AGN
classifications and (b) identify any broadening or shifting of the
\oiii\ emission line.  A detailed description of our multi-component
fitting routine is given in appendix A.  To summarise, up to two
gaussians are fit to each of the forbidden $[$O~{\sc
  iii}$]\lambda4959$, \oiii, $[$N~{\sc ii}$]\lambda6548$ and \nii\
emission lines (to account for any line asymmetries and/or broad
components) and up to three gaussians to each of the permitted \ha\
and \hb\ lines (up to two gaussians to account for the narrow
component of these permitted lines and one gaussian to account for the
broad component in the case of type 1 AGNs).  The two gaussians used
to model the narrow component of these permitted lines are fixed to
have the same profile as the forbidden $[$O~{\sc iii}$]$ and $[$N~{\sc
  ii}$]$ lines.

\subsection{Source Classification}
\label{Classification}
We used the results from our multi-component line fitting routine to
refine the classifications of the AGNs in our sample beyond that which
is capable using the single component fits performed by the standard
SDSS pipeline.  In what follows, we outline how we separated our
sample into broad and narrow line Seyfert 1s (hereafter, BLS1, NLS1
respectively, and referred to collectively as Type 1 AGNs), Type 2
AGNs and those sources that, following analysis with our
multi-component fitting routine, were identified as being non-AGN
dominated (and excluded from further consideration).

\subsubsection{Type 1 AGNs (BLS1s and NLS1s)}
For each source we used the FWHM of the broadest \ha\ component to
identify Type 1 AGNs and distinguish between BLS1s and NLS1s.  By
using \ha\ rather than \hb\ as the discriminator, we will avoid
classifying intermediate type AGNs (e.g., 1.8, 1.9;
\citealt{Osterbrock81}) as NLS1s since, by definition, intermediate
types show evidence of a broad component in the \ha\ line (but not
necessarily in the \hb\ line).  A source was classified as a Type 1
AGN if an extra gaussian (i.e., beyond that needed to fit the
forbidden lines) provided a significantly better fit (at $>$99 per
cent confidence) to the \ha\ line, and that gaussian constitutes at
least 50 per cent of the total emission line flux and has
\fha$>600$~\kms\ (10,548 sources).  The $>50$ per cent cut excludes
995 AGNs that our fitting routine measures \fha$>600$~\kms, 717 (i.e.,
$\sim$72 per cent) of which are subsequently classified as type 2
AGNs.  These 995 AGNs represent $<10$ per cent of either Type 1s or
Type 2s and, subsequently, their reclassification has little impact on
our results yet ensures that we only include genuine type 1 AGNs in
that group.

As per the classical, yet arbitrary, definition of \cite{Goodrich89},
all of those Type 1 AGNs whose broadest \ha\ component has a
\fha$>2000$~\kms\ were sub-classified as BLS1s (9,455 sources).  The
remaining Type 1 AGNs (1,093 sources) were classed as NLS1s (of which
1,024, or $\sim$93 per cent, have broad \ha\ components that satisfy
1000~\kms$<$\fha$<$2000~\kms).

\subsubsection{Type 2 AGNs}
We used the \cite{Kauffmann03} emission line (BPT) diagnostics to
identify Type 2 AGNs in the remaining sample (i.e., those that have
not already been identified as Type 1 AGNs).  Where two gaussians are
fit to the \oiiis\ and \niis\ lines, we use the combined flux of these
components.  In such cases, we use the combined flux of the two narrow
components of the \hb\ and \ha\ lines.  Similarly, when only one
gaussian is needed to fit the \oiiis\ and \niis\ lines, we only use
the flux of the single narrowest gaussian of the \ha\ and \hb\ lines
(i.e., if present, we always exclude any \ha\ and \hb\ broad
components when calculating their narrow-line flux).  In this way, we
classify 13,716 Type 2 AGNs.

\subsubsection{Non-AGN Dominated}
Any narrow line objects (i.e., non-Type 1 AGNs) which, after being
analysed by our multi-component fitting routine, were not classified
as Type 2 AGNs were removed from the sample (1,406 sources), leaving
24,264 confirmed AGNs (of either Type 1 or Type 2).  After analyses
with our emission-line fitting routine, these rejected sources were
found to predominantly lie in the LINER region of the BPT diagram and
have \fha $>600$~\kms\ as measured by the SDSS pipeline.

\subsection{1.4~GHz Radio Coverage}
One of the key aims of this study is to determine how the kinematics
of the \oiii\ emitting gas relates to the radio properties of the AGN.
To quantify the radio properties of our sample we performed a
three-way cross matching between our optical (SDSS) catalogue and the
NVSS and FIRST radio catalogues, largely following the procedure
outlined in \cite{Best05}.  As such, we only use radio fluxes from the
NVSS catalogue, but use the higher-spatial resolution information
provided by the FIRST survey to ensure a low level of contamination
within our optical-radio matched catalogue.  Contrary to
\cite{Best05}, the goal of this study is not to obtain a highly
complete sample of radio sources, but instead to measure the radio
properties of optically selected AGNs.  As such, we considered all
NVSS sources with 1.4~GHz fluxes $>2$~mJy; this lower limit is a
factor of 2.5 below that used by \cite{Best05}.  However, we note that
the integrated 1.4~GHz flux densities of all the sources in our
catalogue have signal-to-noise ratios $>3$.  For those sources not
detected by the NVSS, we assumed an integrated 1.4~GHz flux density
upper limit of 2~mJy.

Of the 10,548 Type 1 and 13,716 Type 2 AGNs in our SDSS sample, 535
(i.e., $\sim$5 per cent) and 1453 (i.e., $\sim$11 per cent),
respectively, are found to have radio counterparts (representing
$\sim$8 per cent of all AGNs in our sample).  The discrepancy between
the fraction of Type 1 AGNs with radio detections compared to that of
Type 2 is likely to be due to the redshift distributions of these
classes of galaxies in our sample.  Since they are brighter at optical
wavelengths, Type 1 AGNs are more likely to satisfy the apparent
magnitude cut for spectral follow-up by the SDSS.  The higher average
redshift of the Type 1 AGNs means a smaller fraction of them are
detected in the NVSS radio survey (which has roughly a constant flux
limit).  

We assume a radio spectral index of $\alpha=0.8$ when calculating the
rest-frame 1.4~GHz radio luminosity, where
$F_{\nu}\propto\nu^{-\alpha}$ (\citealt{Ibar10}).

\subsection{Spectral stacks and average \oiii\ profiles}
In addition to our fits to individual spectra we also rely heavily on
spectral stacking to derive the average \oiii\ profiles of
optically-selected AGNs.  By binning the AGN sample in terms of
luminosity, Eddington ratio and radio properties we can use these
stacks to isolate general trends in the \oiii\ profile with respect to
these parameters.  The benefit of spectral stacking is that it is
less susceptible to systematic biases than results derived solely from
individual fits (e.g., the low S/N of fainter AGNs affecting our
ability to detect broadened line components) and, by considering
relative changes in the average profile of the \oiii\ line, to
measures of the systematic velocity of the AGN.  In \S\ref{Individual}
we use the results from our spectral stacks to motivate further
analyses using the individual AGN spectra.

The spectral stacks were built by coadding the continuum-subtracted
spectra provided by the SDSS database.  Each spectrum was shifted into
the rest frame using the \oiii\ redshift derived from our fits
(assuming that the narrowest \oiii\ component lies at the systematic
redshift).  We note that the redshift derived from the \oiii\ line is
in very good agreement with the redshift provided by the SDSS database
-- a gaussian fit to the $\Delta z=z_{\rm SDSS}-z_{\rm [O~III]}$
distribution reveals an average offset of $\Delta
z=(0.9\pm2.1)\times10^{-4}$ (i.e., $26\pm63$~\kms); indeed, our tests
show there is no significant impact in our results if we adopt the
SDSS-reported redshifts instead. We take a bootstrapping approach
  to calculate the error on the average emission line profiles.  We
  select, at random, a third of the sample in each of our bins and
  stack their spectra.  We repeat this process 1000 times and
  calculate the standard deviation at each wavelength of the resulting
  distribution of stacked spectra.  This standard deviation is then
  adopted as the error on the stacked line profiles.

When stacking as a function of radio properties (i.e., \lrad\ and
\rrad), we only include radio-undetected sources when (a) their upper
limits are consistent with the upper boundary of the bin and (b) there
is no lower boundary on the bin (e.g., \lrad$<10^{22}$~\whz\ and
\rrad$<10^{-1}$).

\section{Results from spectral stacking}
\begin{figure}
\begin{center}
	\includegraphics[width=8.3cm,height=8.6cm]{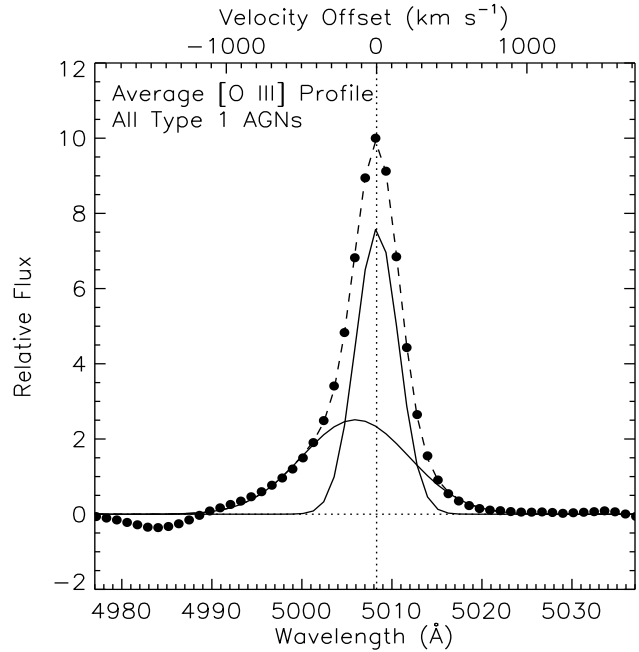}
\end{center}
\caption{Average \oiii\ profile of Type 1 AGNs in our sample derived
  from spectral stacking.  This profile shows a strong blue wing
  (blueward asymmetry) that is well modelled by a broad (i.e.,
  FWHM$=851$~\kms) second gaussian component that is offset from the
  narrower component (i.e., FWHM$=335$~\kms) by $148$~\kms.  The fit
  produced by combining these two components is shown by the dashed
  line.}
\label{Coadd_Sy1}
\end{figure}

\label{Results}
In our analyses we explore the kinematics of the AGNs' narrow line
regions as portrayed by the profiles of the forbidden \oiii\ emission
lines.  Using stacked spectra, we search for general trends in the
profiles of the \oiii\ lines with respect to various key AGN
parameters (specifically, bolometric luminosity, Eddington ratio,
radio loudness, radio luminosity and AGN orientation).  In
\S\ref{Individual} we use the results from the emission line fitting
routine described in \S\ref{FittingRoutine} (see also appendix
\ref{linefit}) to estimate the proportions of optically selected AGNs
that have either very broad or highly shifted \oiii\ lines or
components.

We initially focus our analyses on Type 1 AGNs. Assuming outflows are
driven radially outwards perpendicular to the plane of the dusty
torus/accretion disk, then from AGN unification models (e.g.,
\citealt{Antonucci93}) any outflowing material driven by the AGN
(either by jets or radiatively driven winds) should show a stronger
blueshift (asymmetry) in the spectra of Type 1 AGNs (compared to Type
2s) as the outflow will be directed more along our line of sight. As
such, if the AGNs are indeed driving outflows, and these outflows emit
\oiii, we should expect Type 1 AGNs to show stronger \oiii\ line
shifts. In \S\ref{Results:Obscuration} we compare the results derived
from our Type 1 population with the Type 2 AGNs in our sample to
explore the influence of AGN orientation on the observed emission line
profiles and measured gas kinematics.

\subsection{The average \oiii\ profile of Type 1 AGNs}
\label{Average_Type1}
In \fig{Coadd_Sy1} we present the average \oiii\ profile derived from
stacking all Type 1 AGNs in our sample.  This profile displays a
prominent blue wing and is well fit using two gaussian components; a
narrow (i.e., FWHM$=335$~\kms) component and a broader (i.e.,
FWHM$=851$~\kms) component that is blueshifted by 148~\kms\ with
respect to the narrow component.  The flux contained within the
broader component represents $\approx$45 per cent of the total line flux.  

\begin{figure*}
\begin{center}
	\includegraphics[width=16cm,height=16cm]{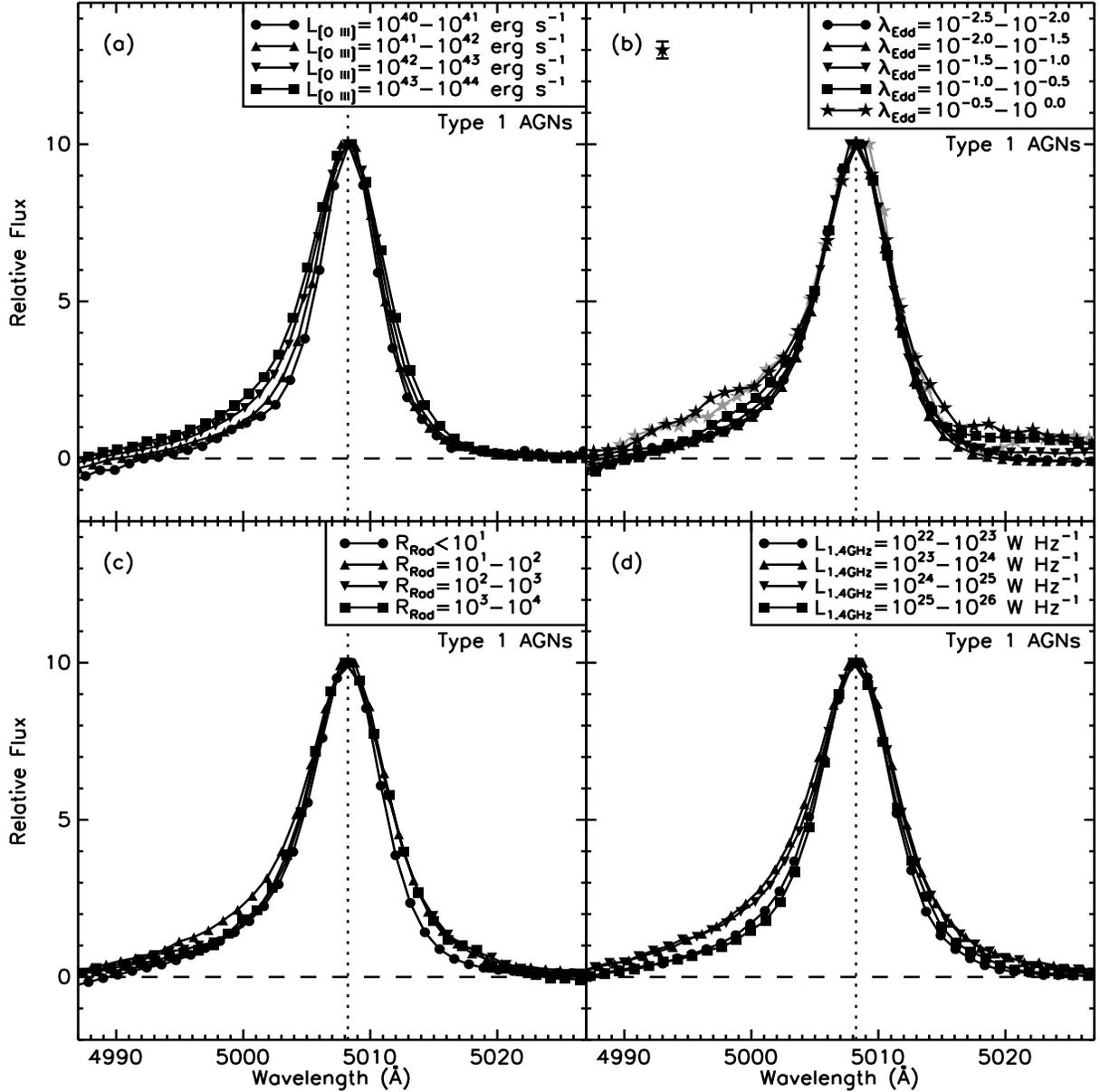}
\end{center}
\caption{Average \oiii\ profiles of Type 1 AGNs binned according to
  (a) [O~III] luminosity, (b) Eddington ratio, (c) radio loudness and
  (d) radio luminosity.  The grey curve in plot (b) shows the average
  profile of the highest Eddington ratio AGNs after excluding those
  spectra that show prominent Fe~{\sc ii} emission that can
  contaminate \oiii.  For presentation purposes we do not show the
  errors on the stacked profiles which, with the exception of the
  highest Eddington ratio bin in panel (b) (see representative error
  bar), are smaller that the points.}
\label{stack_page}
\end{figure*}

In the following subsection we consider whether the profile of the
\oiii\ line changes as a function of bolometric luminosity (\lagn),
Eddington ratio (\redd), radio loudness (\rrad), radio luminosity
(\lrad) and orientation.  We do this by binning the sample in terms of
these properties and use spectral stacking to derive the average
\oiii\ emission line profiles of the AGNs in each bin.

\subsection{Correlations between the average \oiii\ profile and key
  AGN properties}
\label{Correlations}
\subsubsection{$[{\rm O}$\ {\sc iii}$]\lambda5007$ luminosity}
\label{Results:Bolometric} 

If outflows from AGNs are driven by radiative forces, and these
outflows manifest themselves as blueshifted \oiii\ wings, then we may
expect the observed properties of these wings to change as a function
of the AGN luminosity.  The \oiii\ luminosity (hereafter, \loiii) is
thought to be a proxy measure of the bolometric AGN luminosity
(e.g. \citealt{Maiolino95, Risaliti99, Bassani99, Heckman05}), so here
we investigate the relationship between \oiii\ profile and \loiii.

In the top-left panel of \fig{stack_page} we present the average
\oiii\ profiles of Type 1 AGNs in our sample divided into bins of
log(\loiii), with bin intervals of $\Delta$log(\loiii)=1 over
\loiii$=10^{40}-10^{44}$~\ergs.  We calculate \loiii\ using the
results from our multi-component fitting routine, assuming the
redshift calculated from the \oiii\ line; when two \oiii\ components
are present, \loiii\ is the combined luminosity of both components,
although we note that our results remain unchanged when only the
luminosity of the narrow \oiii\ component is used.  We use the Balmer
decrement (using the ratio of the narrow \ha\ and \hb\ components
calculated from our fits) and assume the \cite{Cardelli89} reddening
curve to correct for the effects of reddening on \loiii.

The stacked \oiii\ emission line profiles of Type 1 AGNs display
increasingly prominent blue wings with increasing \loiii.  However,
before interpreting this result as evidence that radiative forces are
largely responsible for the kinematics of the \oiii\ emitting gas, it
is important to account for the known positive correlation between
\loiii\ and AGN radio power (e.g., \citealt{Baum89, Rawlings89,
  Zirbel95, Tadhunter98, Wills04, deVries07}).  For this, we perform the same
experiment as described above (i.e., stacking as a function of \loiii)
however this time controlling for radio luminosity to ensure that the
median radio luminosities of the stacks are consistent to within a
factor of 3 of each other (see \Tab{Comparison}).  We have also
attempted to keep the number of sources in each stack roughly the same
by randomly selecting, without replacement, sources from each bin.
The numbers in each stack are therefore constrained by two effects:
(i) the number of sources in the bin containing the smallest number of
AGNs and (ii) the ability to select sources from each bin with radio
luminosities that are consistent to within a factor of 3.  The
combination of these two effects is why the numbers of stacked sources
are not {\it exactly} the same in each of the bins.
\begin{figure*}
\begin{center}
	\includegraphics[width=16cm,height=16cm]{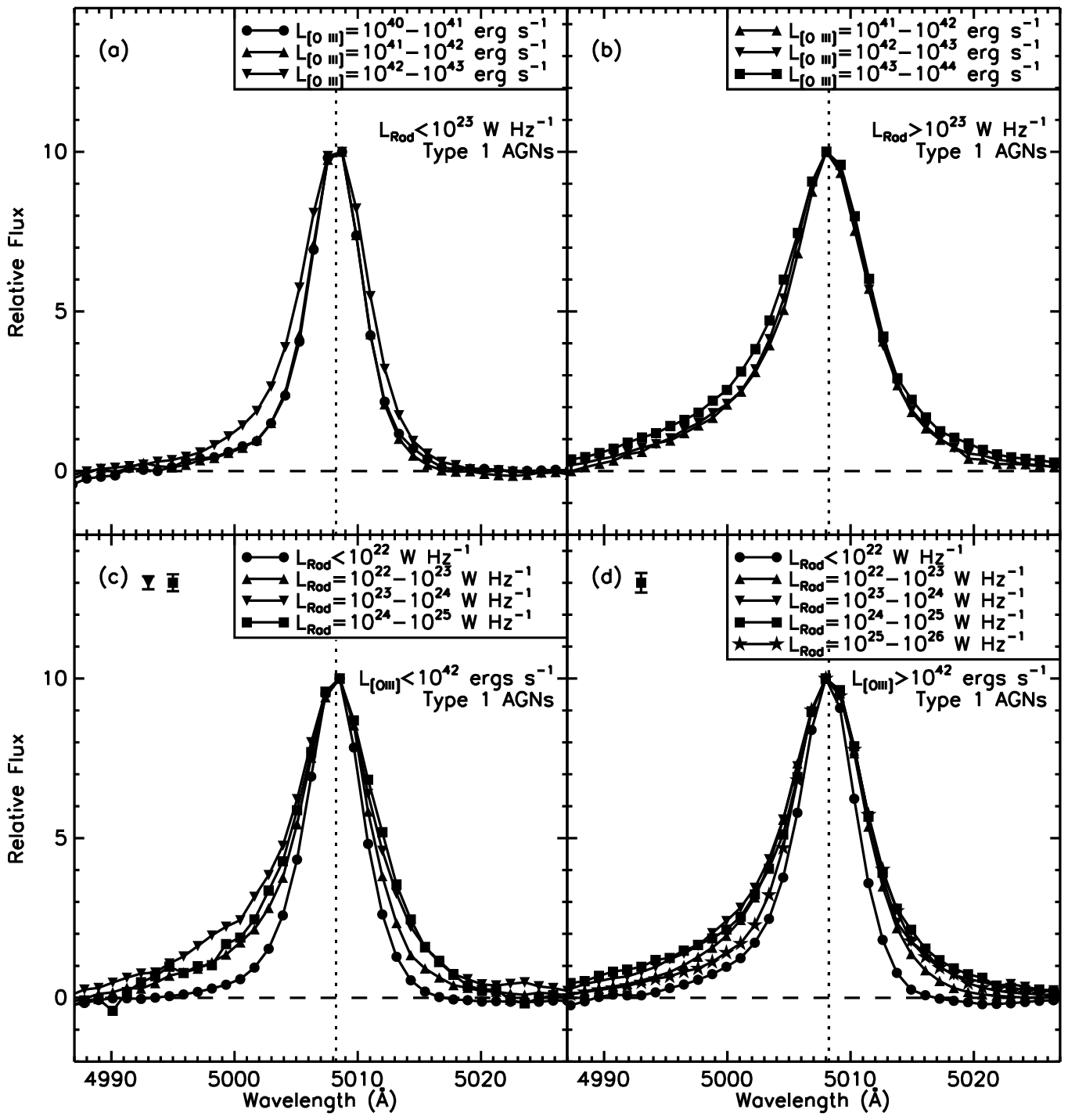}
\end{center}
\caption{Average \oiii\ profiles of Type 1 AGNs, accounting for the
  correlation between \loiii\ and \lrad\ in an attempt to isolate
  which has the greater affect on the line profiles.  Panels (a) and
  (b) show the results of stacking as a function of \loiii, while
  keeping \lrad\ roughly fixed (to $<10^{23}$~\whz\ and
  $10^{24}$~\whz, respectively).  Alternatively, panels (c) and (d)
  show the results of stacking as a function of \lrad\, keeping
  \loiii\ roughly fixed (to $\sim5\times10^{41}$~\ergs\ and
  $\sim2\times10^{42}$~\ergs, respectively).  The greatest change in
  the profiles of the \oiii\ line is seen when \lrad\ varied.  As with
  fig. \protect \ref{stack_page} we show representative error bars when
  they are larger than the plotted points (see panels (c) and (d).}
\label{comp_stack_page}
\end{figure*}

The results of this experiment are shown in the top two panels of
\fig{comp_stack_page}.  Here, we have divided the sample into radio
weak (i.e., \lrad$<10^{23}$~\whz, including any upper limits that
also satisfy this criterion; panel {\it a}) and radio powerful (i.e.,
\lrad$>10^{23}$~\whz, excluding all upper limits; panel {\it b}).
When the stacks are controlled for radio luminosity, the change in
average \oiii\ profile as a function of \loiii\ is much less
pronounced in both the radio weak and radio strong
samples.\footnote{We note that we do not calculate the average profile
  for the highest \loiii\ bin (i.e., \loiii$=10^{43}-10^{44}$~\ergs)
  of the radio weak AGNs, nor the lowest \loiii\ bin (i.e.,
  \loiii$=10^{40}-10^{41}$~\ergs) of the radio strong AGNs, as there
  are too few sources to provide reliable averages - a consequence of
  the correlation between \loiii\ and radio power.}  We do note,
however, that a slight increase in the prominence of the \oiii\ blue
wings with increasing \loiii\ still remains in both the radio strong
and weak subsamples.  However, comparing the top two panels of
\fig{comp_stack_page} reveals a striking difference in the average
profiles of radio weak and radio strong AGNs, with all the radio
strong stacks having broader \oiii\ profiles than their equivalent
radio weak stack.  We explore this further in \S\ref{Results:Radio}.

\begin{figure*}
\begin{center}
	\includegraphics[width=16cm, height=8.6cm]{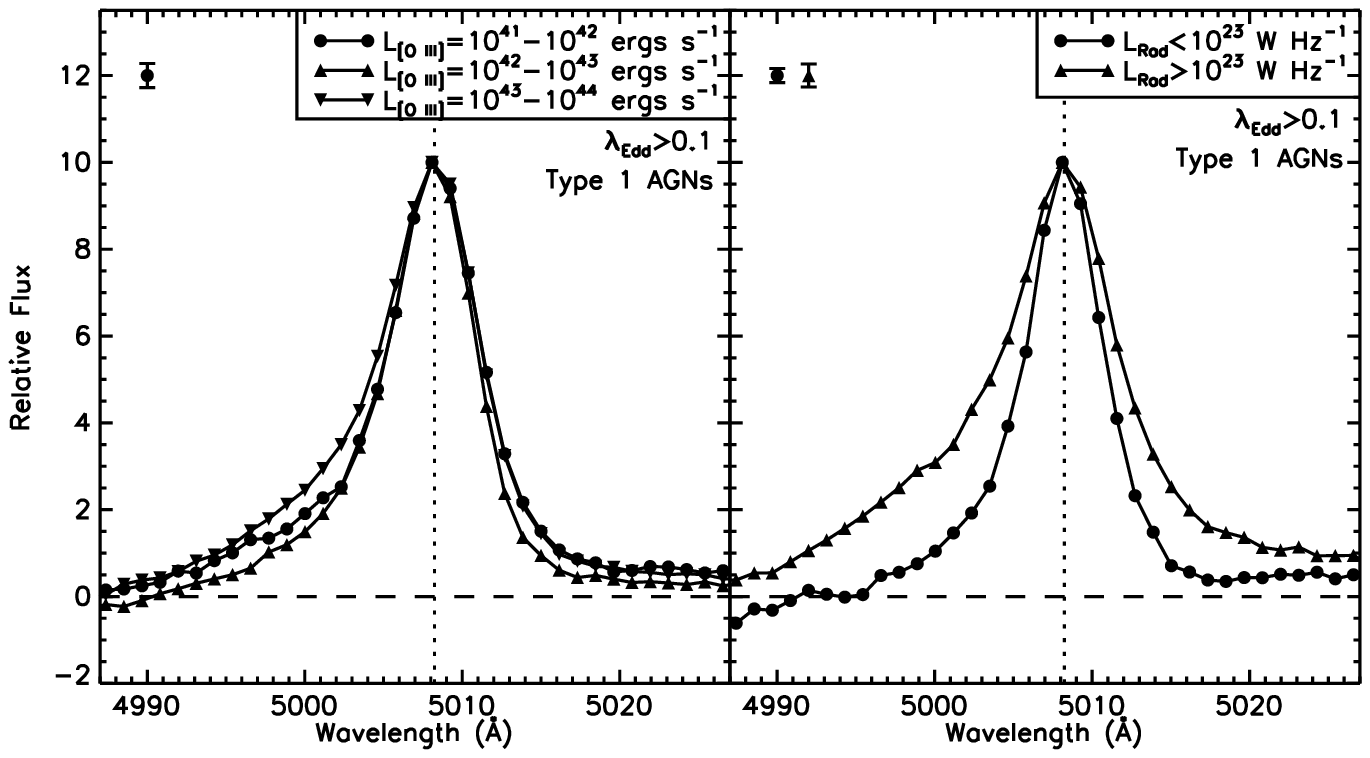}
\end{center}
\caption{Average \oiii\ profiles of high Eddington-ratio (\redd$>0.1$)
  Type 1 AGNs, split as a function of \oiii\ luminosity (left) and
  radio luminosity (right).  Similar to Fig. \protect
  \ref{comp_stack_page}, we have stacked in different \loiii\ and
  \lrad\ bins (see keys), matching \redd\ in each bin to keep the
  average \redd\ roughly the same between the two bins.  As with
  fig. \protect \ref{stack_page} we show representative error bars
  when they are larger than the plotted points.}
\label{comp_edd}
\end{figure*}

\begin{figure*}
\begin{center}
	\includegraphics[width=16.cm, height=9cm]{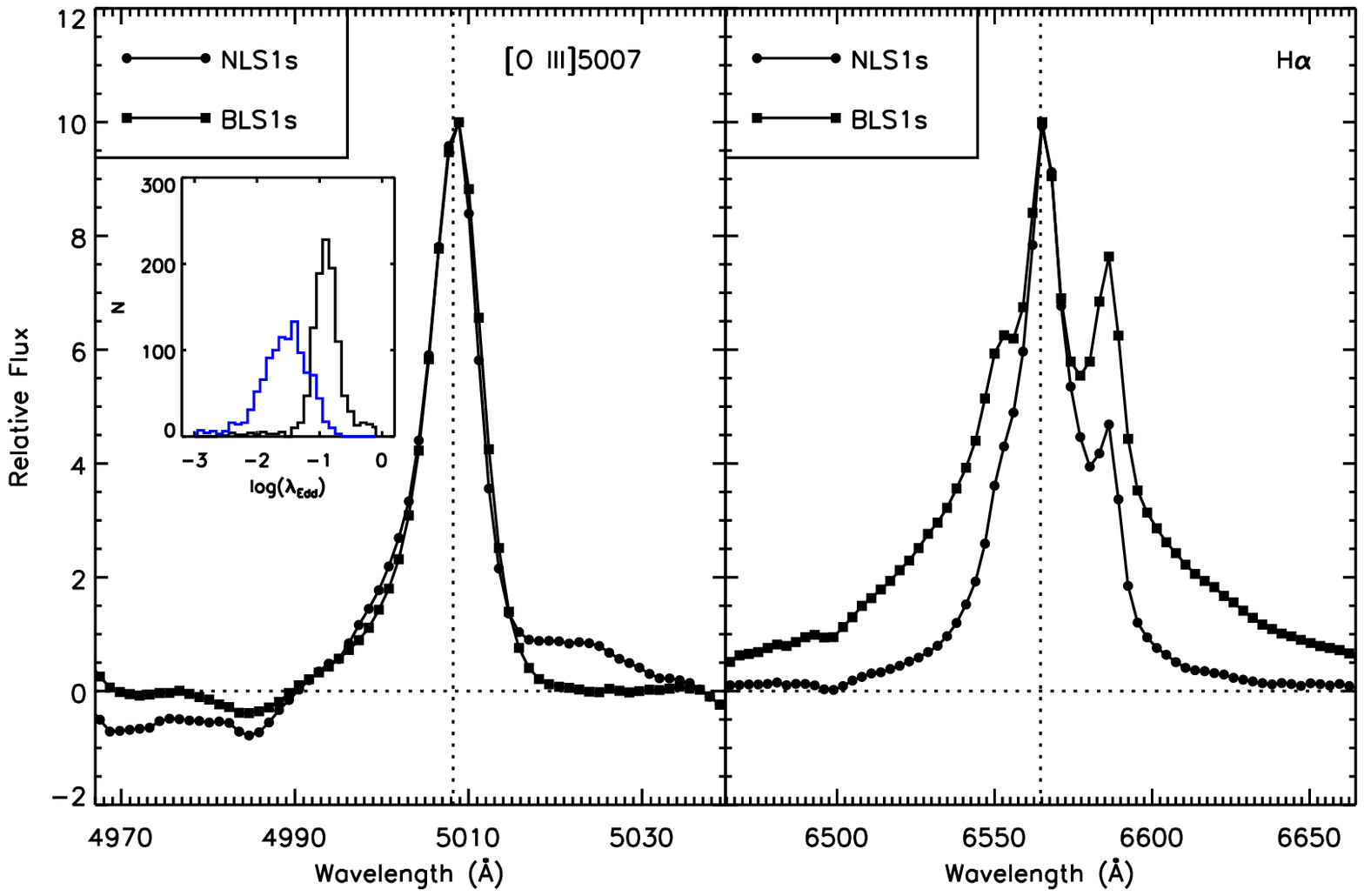}
\end{center}
\caption{Average \oiii\ profile (left) and \ha\ profile (right) of the
  BLS1 and NLS1 AGNs in our sample.  Inset we show the
    distribution of Eddington ratios for the NLS1s (black histogram)
    and the BLS1s (blue histogram) in our sample, indicating that the
    NLS1s have systematically higher Eddington ratios. We find no
  evidence to suggest that, on average, NLS1s have \oiii\ lines that
  are either broader or more blueshifted than their BLS1
  counterparts.}
\label{NLS1s}
\end{figure*}

\subsubsection{Eddington ratio}
\label{Results:Eddington}
Previous studies of the \oiii\ line profile have concluded that the
Eddington ratio (i.e., $\lambda_{\rm Edd}=L_{\rm AGN}/L_{\rm Edd}$,
where $L_{\rm Edd}$ is the Eddington luminosity of the central SMBH)
of the AGN plays a key role in dictating both the FWHM and blueshift
of the \oiii\ line, including whether it displays a strong blue wing
(e.g., \citealt{Greene05}).  Here, we consider whether our analyses
confirm that the \oiii\ profile is correlated with the Eddington ratio
of the AGN.

To estimate the Eddington luminosity of the AGN we use the SMBH mass
calculated from the FWHM and luminosity of the \ha\ emission line (see
eqn. 6 of \citealt{Greene05b}).  We only calculate $L_{\rm Edd}$ for
Type 1 AGNs and use only the FWHM of the broadest \ha\ gaussian.
Here, the bolometric luminosity (i.e., $L_{\rm AGN}$) is estimated
from the luminosity of the \ha\ line using a combination of the
$L_{H\alpha}$-$L_{5100}$ and $L_{5100}$-$L_{AGN}$ correction factors
described in \cite{Greene05} and \cite{Netzer07}, respectively.  These
combine to give an expression for the Eddington ratio of:

\begin{equation}
\lambda_{\rm Edd} = \frac{L_{\rm AGN}}{L_{\rm Edd}} \approx 
0.43
f_{10}
\left( \frac{L_{\rm H\alpha}}{10^{42}~{\rm ergs~s^{-1}}} \right)^{0.31}
\left( \frac{\rm FWHM_{H\alpha}}{10^{3}~{\rm km~s^{-1}}} \right)^{-2}
\end{equation}

\noindent
where $f_{10}$ is the dimensionless $L_{5100}$-$L_{AGN}$ bolometric
correction factor (see \citealt{Netzer07}) in units of 10.  We assume
$f_{10}~=~1$ throughout, although note that this has no effect on our
final results as we are only concerned with relative changes in the
\oiii\ profile.

\begin{table} 
\begin{center}{
\caption{Sample properties after accounting for the correlation
  between \loiii\ and \lrad.}
\label{Comparison}
\input{./Table1.tex} }\end{center}{\sc Notes}: The first six
rows are the samples that have been selected to have roughly constant
\lrad\ but increasing \loiii. The lower nine rows are those samples that
have been selected such that \loiii\ remains roughly constant, but
\lrad\ increases. (1) Range of \lrad; (2) range of \loiii; (3) the
number of sources in each subsample; (4) the median radio luminosity;
(5) the median \oiii.
\end{table}

The average profiles of the Type 1 AGNs in our sample, stacked in
terms of \redd\ are presented in the top-right panel of
\fig{stack_page}.  We find no significant difference in the average
\oiii\ profiles over the range \redd$=3\times10^{-3}-3\times10^{-2}$,
while the \oiii\ profiles of \redd$=3\times10^{-2}-1$ AGNs display
increasingly prominent blue wings with increasing \redd.  The average
spectrum of AGNs in the highest \redd\ bin that we consider (i.e.,
\redd$=0.3-1$) shows evidence of contamination from Fe {\sc ii}
(beyond, e.g., $\sim5020$\AA) that could alter the apparent profile of
the \oiii\ line.\footnote{There is a known correlation between the
  equivalent widths of the Fe {\sc ii} complex and \redd, see
  \cite{Boroson92,Boroson02}.} When we remove those AGNs in the
highest \redd\ bin that show evidence of having strong Fe~{\sc ii}
emission at other wavelengths in their spectrum, the resulting stacked
\oiii\ profile remains largely unchanged.  As with our stacks as a
function of \loiii, we must use caution before directly associating
this broadening with increased \redd, as there may be other, more
fundamental, drivers.  To test for this, we performed a similar
experiment as outlined at the end of \S\ref{Results:Bolometric};
splitting the high \redd\ AGNs into bins of \loiii\ and \lrad while
controlling for \redd.  Unfortunately, there are too few AGNs in our
highest \redd\ bin (i.e., \redd$>0.3$) to perform a meaningful
experiment (there are only six Type 1 AGNs in our sample with
\redd$>0.3$ and \lrad$>10^{23}$~\whz), so we resort to \redd$>0.1$).
The results of this experiment is shown in \fig{comp_edd}.  As with
the full Type 1 AGN sample, we see a broadening of the \oiii\ line
with \loiii\ for \redd$>0.1$ AGNs, but note that these stacks have not
been controlled for \lrad, so some of this broadening may be related
to the correlation between \loiii\ and \lrad.  Indeed, we see a
striking difference between the \oiii\ profiles of high and low \lrad\
AGNs with high \redd, suggesting that even in these high \redd\ AGNs,
mechanical processes are playing an important role in the disturbing
the \oiii\ emitting gas.  However, we stress that this only applies to
\redd$>0.1$ AGNs, as the small numbers of \redd$>0.3$ AGNs in our
sample means we cannot perform this experiment in the highest \redd\
bin where we see the strongest blue wing.

\subsubsection{Radio loudness and luminosity}
\label{Results:Radio}
In addition to radiative forces, it is also thought that radio jets
(either compact or extended) may play a significant role in disturbing
the interstellar medium around an AGN, potentially inducing feedback
(e.g., \citealt{Whittle92, Nelson96, Tadhunter01, Holt03, Holt08}).
Indeed, \cite{Nesvadba08} have suggested that the spatially resolved
broadening of the \oiii\ lines in the spectra of high redshift radio
galaxies are the result of interactions between a radio jet and the
NLR.  If this the case, then we may expect the profiles of the \oiii\
line to be related to the radio properties of the AGNs.

\begin{figure}
\begin{center}
  \includegraphics[width=8.3cm, height=8.6cm]{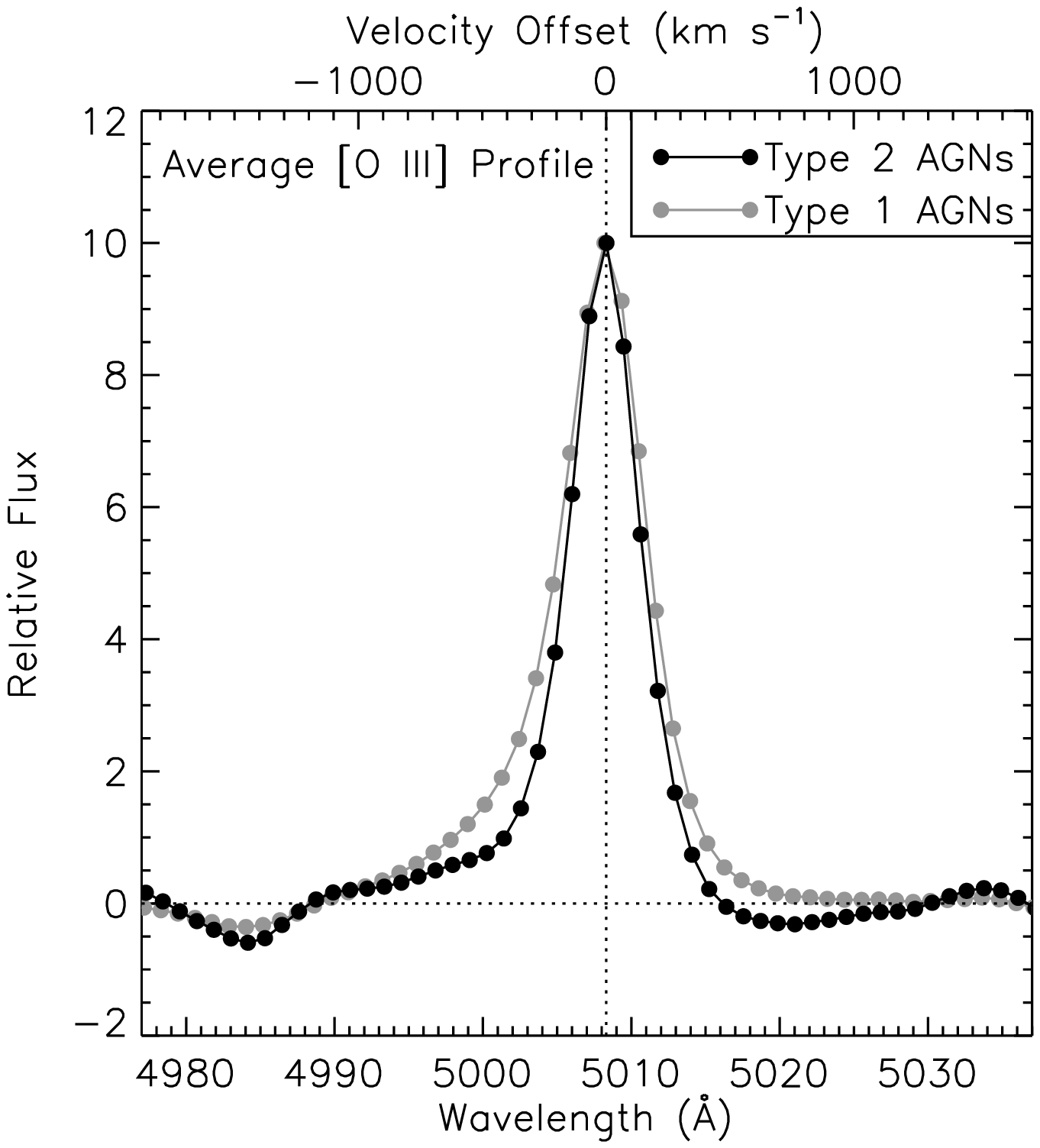}
\end{center}
\caption{Average \oiii\ profile of the Type 2 AGNs in our sample,
  derived from spectral stacking.  This average profile shows less
  prominent blue wings compared to the average Type 1 profile (shown
  in grey).  The absorption either side of the \oiii\ line is
  attributable to the stellar features in the host galaxy component
  which are stronger in Type 2 AGNs due to the central engine
  continuum being obscured from view.}
\label{Coadd_Sy2}
\end{figure}

In the lower two panels of \fig{stack_page} we present the average
\oiii\ profiles of the Type 1 AGNs in our sample, stacked according to
radio loudness (defined here as $R_{\rm Rad}=F_{\rm 1.4~GHz}/F_{\rm
  B}$; where $F_{\rm 1.4~GHz}$ and $F_{\rm B}$ are the flux densities
measured in the 1.4~GHz band and the rest-frame optical B-band,
respectively) and radio luminosity (at 1.4~GHz; hereafter,
\lrad).\footnote{B-band fluxes were calculated by passing the
  rest-frame SDSS spectra through a synthetic Buser B-band response
  filter.}  The average \oiii\ profiles show evidence of only a modest
change across over three orders of magnitude in \rrad\ (i.e.,
\rrad$<10$ to $10^{3}-10^{4}$).  As an aside, we find that the average
\oiii\ lines of radio detected (irrespective of loudness) AGNs have
profiles similar to those AGNs with the highest \loiii\ values (i.e.,
\loiii$>10^{42}$~\ergs).  However, this result is not surprising when
we consider that the radio detected AGNs in our sample typically have
high \oiii\ luminosities.  For example, 75 per cent (i.e., 654/871) of
radio detected Type 1 AGNs have \loiii$>10^{42}$~\ergs).

We also investigated whether any trend exists between the average
\oiii\ profile and radio luminosity (as opposed to radio loudness).
From our spectral stacks we find that there is a non-monotonic change
in the average \oiii\ profiles over the four orders-of-magnitude range
in \lrad\ spanned by our sample; the \lrad$=10^{23}-10^{24}$~\whz\ and
\lrad$=10^{24}-10^{25}$~\whz\ stacks display a marginally stronger
blue wing compared to the \lrad$=10^{22}-10^{23}$~\whz\ and
\lrad$=10^{25}-10^{26}$~\whz\ stacks.

As outlined in \S\ref{Results:Bolometric}, it is important that we
consider the influence that the correlation between \loiii\ and \lrad\
may have on our results.  For this, we employ a similar method as that
used in \S\ref{Results:Bolometric}.  This time, however, we control
for \loiii\ while stacking as a function of \lrad, ensuring that the
median \loiii\ of the stacked sources are consistent to within a
factor of 3.  The results of this experiment are shown in the lower
two panels of \Fig{comp_stack_page}.  Here, we have separated the
sample into high and low \oiii\ luminosity subsamples (i.e.,
\loiii$>10^{42}$~\ergs\ and \loiii$<10^{42}$~\ergs, respectively).  In
both subsamples we see a clear change in the average \oiii\ line
profiles as a function of \lrad.  We note again, however, that this
increase is not monotonic.  Instead, the \oiii\ profile broadens with
increasing \lrad\ from \lrad$<10^{22}$~\whz\ to
\lrad$=10^{23}-10^{24}$~\whz\ (\lrad$=10^{23}-10^{25}$~\whz\ in the
case of \loiii$>10^{42}$~\ergs\ AGNs), but then decreases in width as
\lrad\ increases further (to \lrad$=10^{24}-10^{25}$\ \whz\ and
\lrad$=10^{25}-10^{26}$\ \whz\ for the \loiii$<10^{42}$~\ergs\ and
\loiii$>10^{42}$~\ergs\ subsamples, respectively).

\subsubsection{The average $[{\rm O}$\ {\sc iii}$]\lambda5007$
  profiles of Narrow-line Seyfert 1s}

\begin{figure*}
\begin{center}
	\includegraphics[width=16.6cm, height=8.2cm]{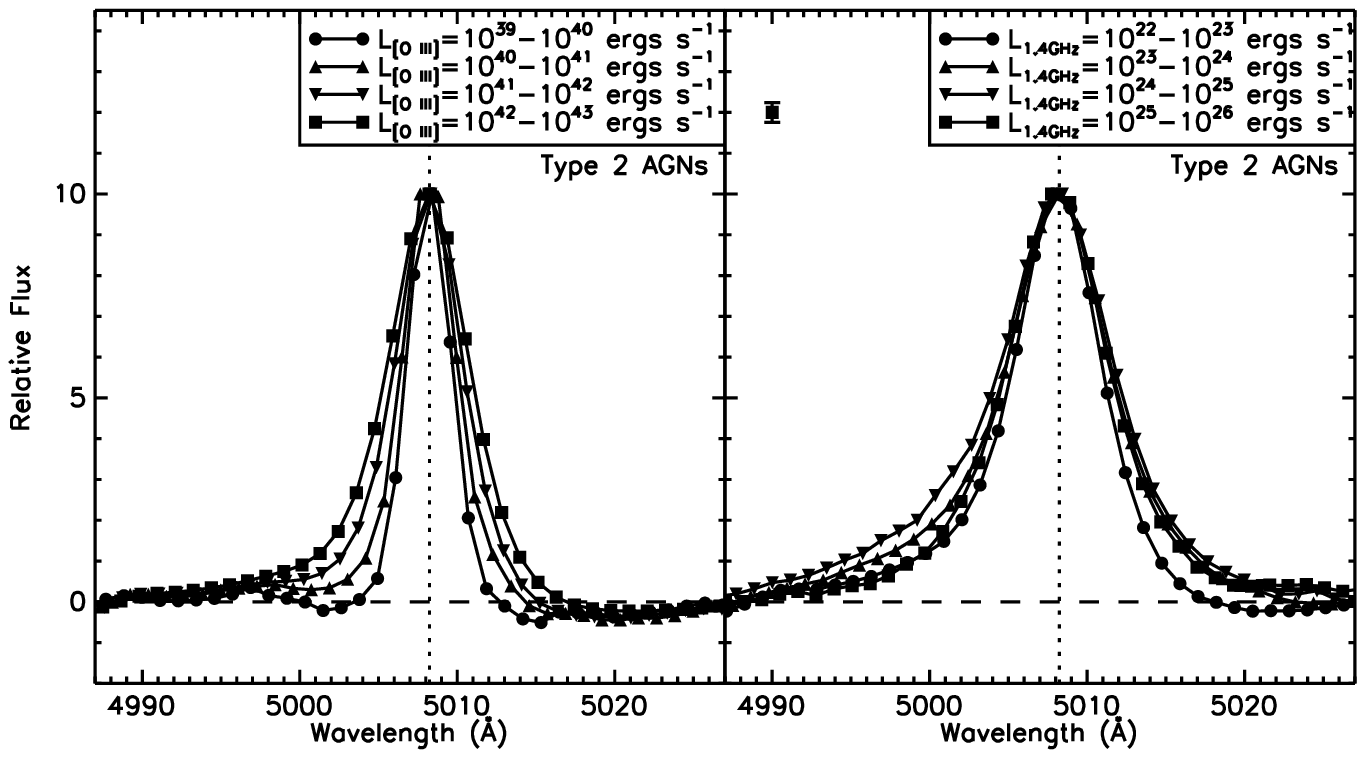}
\end{center}
\caption{Average \oiii\ profiles of Type 2 AGNs binned according to
  \oiii\ luminosity.  As is the case for Type 1 AGNs (see
  \fig{Coadd_Sy1}), we find that the average \oiii\ profile of Type 2
  AGNs increases in width with increasing \oiii\ luminosity.  However,
  the Type 2 profiles show much less prominent blueward asymmetries
  compared to the average \oiii\ profiles of Type 1 AGNs.  As with
  fig. \protect \ref{stack_page} we show representative error bars
  when they are larger than the plotted points.}
\label{sy2_stack}
\end{figure*}

It has previously been suggested that Type 1 AGNs with narrow Balmer
lines (i.e., $<4000$~\kms) and, in particular, NLS1s show a tendency
to have more prominent \oiii\ blue wings or blueshifts compared to
BLS1s (e.g., \citealt{Marziani03, Komossa08}) - a feature previously
attributed to the higher Eddington ratios of NLS1s.  We can test for
this using our sample of Type 1 AGNs.  In \fig{NLS1s} we present the
stacked \oiii\ profile of the NLS1s in our sample compared against
that of BLS1s.  Here, we have stacked the same number of randomly
selected BLS1s as there are NLS1s in our sample (i.e., 1096).  We find
no significant difference between the average \oiii\ profiles of BLS1s
and NLS1s in our sample.  As a check, we also show in this plot the
stacked \ha\ profiles, confirming that our sample of NLS1s do, indeed,
have significantly narrower permitted emission lines (i.e., broad
component \fha\ of 4460~\kms and 1590~\kms; typical for these
respective classes of object).  Furthermore, the NLS1s in our
  sample have systematically higher Eddington ratios than their BLS1
  counterparts, as highlighted in the inset plot showing the
  distribution of \redd\ for the BLS1 and NLS1 samples.  Finally, we
note that the profile of the stacked \oiii\ is largely independent of
our criteria for identifying NLS1s; there is no significant change in
the \oiii\ profile when we change our NLS1 selection criteria to
\fha$<1500$~\kms\ or $<2500$~\kms; in other words the average \oiii\
profile is not dominated by NLS1 AGNs with \fha\ close to the
(arbitrary) boundary between NLS1s and BLS1s.

\subsection{Comparison of \oiii\ profiles of Type 1 \& Type 2 AGNs}
\label{Results:Obscuration}

Up to this point, we have only considered the \oiii\ profiles of Type
1 AGNs.  The observed differences between Type 1 and Type 2 AGNs are
now generally regarded as being the result of obscuration caused by
gas and dust along our line of sight to their central regions (e.g.,
\citealt{Antonucci93}).  It is therefore of interest to consider Type
2 AGNs in order to gain insight into the role that AGN orientation and
obscuration plays in the profiles of the \oiii\ emission lines.

\begin{figure}
\begin{center}
	\includegraphics[width=8cm]{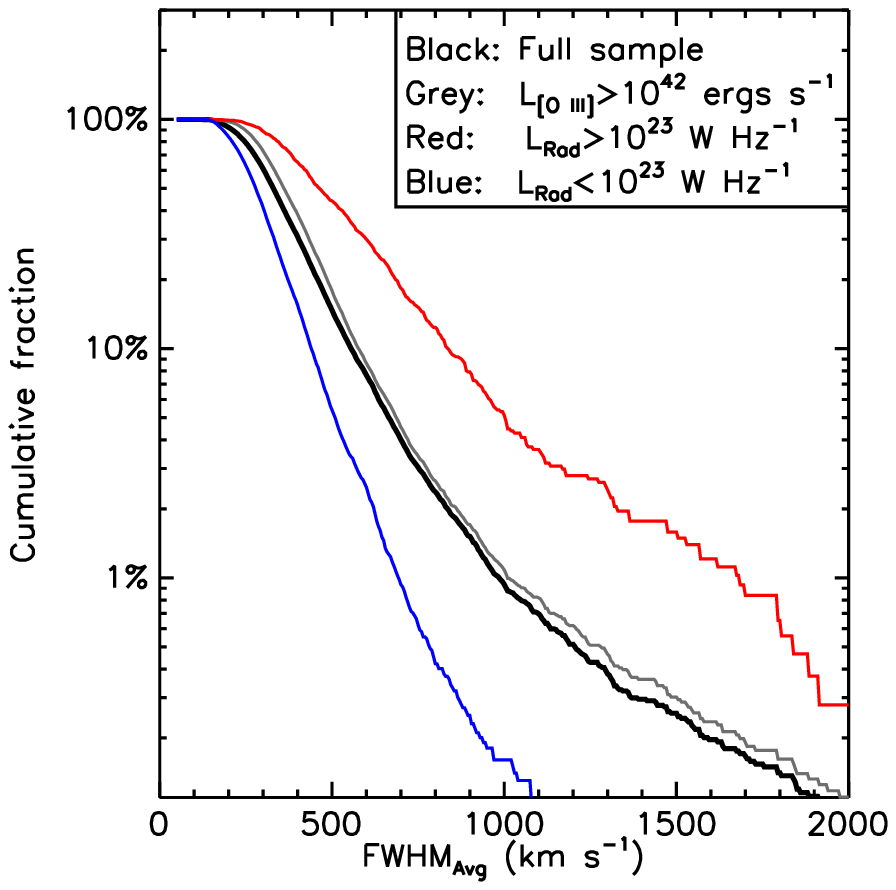}
\end{center}
\caption{The fraction of AGNs in our samples with \fwhma\ greater than
  a given value (shown on the abscissa). \oiii\ lines with \fwhma\
  $>1000$~\kms\ are $\approx$5 times more prevalent among AGNs with
  \lrad$>10^{23}$~\whz\ compared to the overall AGN population.  We
  have included upper limits in the \lrad$<10^{23}$~\whz\ sample when
  they are consistent with this limit.}
\label{fractions}
\end{figure}

In \fig{Coadd_Sy2} we present the average \oiii\ profile derived from
stacking all 13,713 Type 2 AGNs in our sample.  For comparison, we
also include the average \oiii\ profile of the Type 1 AGNs in our
sample (presented previously in \fig{Coadd_Sy1}).  Compared to the
average \oiii\ profile of the Type 1 AGNs, the Type 2s display a less
prominent blue wing, (i.e., is more symmetrical).  Since the \oiii\
line broadens with increasing \loiii\ (but see \S\ref{Results:Radio}),
the narrower Type 2 profile may in part be the result of the lower
average \loiii\ of the Type 2 AGNs in our sample compared to Type 1s
(due to dilution from the strong continuum in the latter; median
\loiii$=6\times10^{41}$~\ergs and $=1.8\times10^{42}$~\ergs,
respectively).  Having said that, as with the stacks of the Type 1
AGNs, the \oiii\ profiles Type 2 AGNs become increasingly broad with
increasing \loiii, but remain more symmetric than the Type 1 stacked
profiles, as recently reported by \cite{Vaona12} (see \fig{sy2_stack},
left-hand panel).  Similarly, we note that the \oiii\ profiles of the
Type 2 AGNs follow the roughly the same trend with \lrad\ as the Type
1s (i.e., increasing in width to \lrad$=10^{23}-10^{24}$~\whz, then
narrowing to higher \lrad) while, again, remaining more symmetrical
(see \fig{sy2_stack}, right-hand panel).  The more symmetrical \oiii\
profiles of Type 2 AGNs are expected if the central engine is driving
an \oiii\ emitting outflow along the axis perpendicular to the plane
of the obscuring torus.  In this model, orientation effects mean that
outflows from type 1 (i.e., unobscured) AGNs will appear more strongly
blueshifted than those from type 2 (i.e., obscured) AGNs.  This is
consistent with \cite{Harrison12}, who reached the same conclusion
using spatially resolved (i.e., integral field unit) spectroscopy.

\section{Results from individual emission line fits}
\label{Individual}
We have shown that the average \oiii\ emission line profile of
optically selected Type 1 AGNs displays a strong blue wing and that,
once the correlation between \loiii\ and \lrad\ has been taken into
account, the width of the average \oiii\ is most strongly related
(though not in a monotonic sense) to the radio luminosity of the AGN.
We see no evidence that the average profile of the \oiii\ line changes
strongly as a function of radio loudness or Eddington ratio, and find
that \loiii\ (assumed to be a proxy for \lagn) has a relatively minor
impact on the average profile (compared to the radio luminosity).
There is no difference in the average \oiii\ profiles of BLS1s and
NLS1s, while Type 2 AGNs have, on average, more symmetric \oiii\
emission lines compared to Type 1s but their widths follow the same
general trends.

In this section, we further explore the results derived from the
spectral stacks by using the results from our fits to the \oiii\
emission line of the individual spectra in our AGN sample.  In doing
so, we will demonstrate that the vast majority of all AGNs have
relatively narrow average \oiii\ profiles, while it is the AGNs with
$10^{23}<$\lrad/\whz$<10^{25}$ that dominate the numbers with the
broadest \oiii\ lines.

\begin{figure*}
\begin{center}
	\includegraphics[width=15cm]{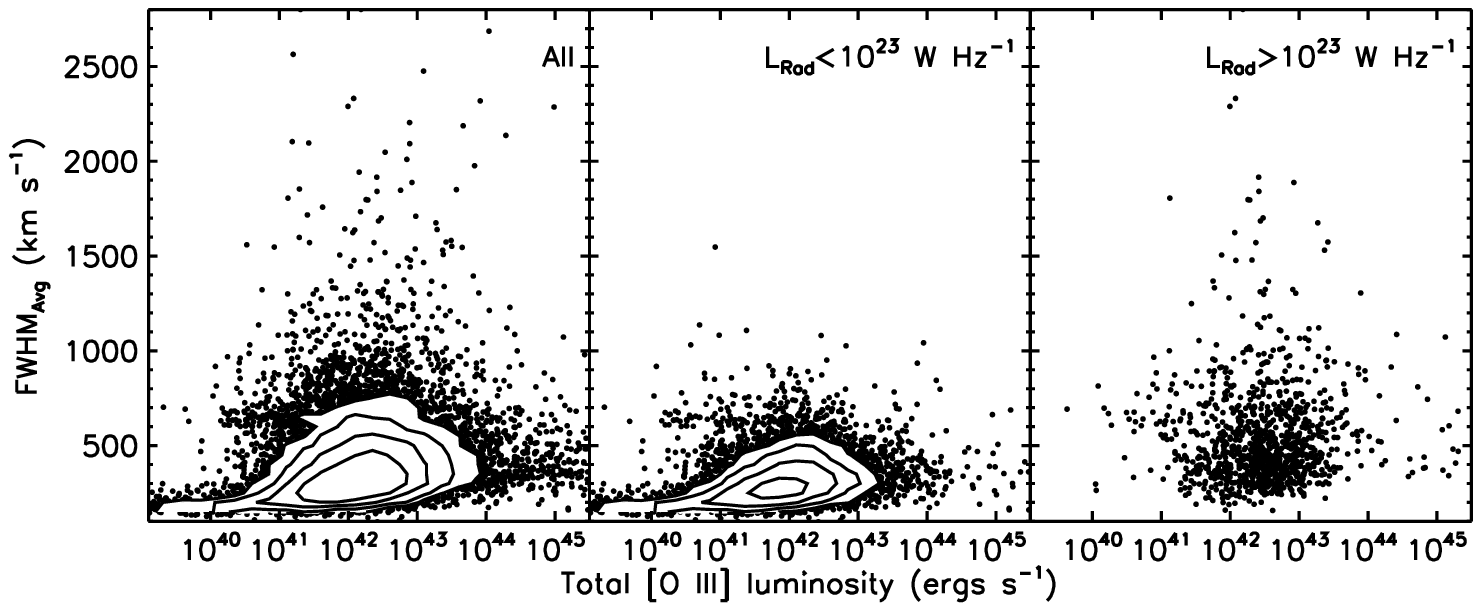}
\end{center}
\caption{The distribution of \fwhma\ (see Eqn. \ref{fwhma}) plotted as
  a function of \loiii\ for all (left-hand panel),
  \lrad$<10^{23}$~\whz\ (middle panel) and \lrad$>10^{23}$~\whz\
  (right-hand panel) AGNs.  The upper envelope of this distribution
  peaks around $10^{42}$~\ergs.  However, the comparing the middle and
  right hand panels reveals that the AGNs with the broadest tend to
  be the more radio luminous of our sample.}
\label{opt_dist_page}
\end{figure*}

We use the results from our multi-component fitting routine to
quantify the proportions of optically selected AGNs with broadened
\oiii\ lines.  Our fitting routine fits up to two gaussian components
to the \oiii\ lines.  However, rather than focus on the properties of
single components, we calculate the flux-weighted average FWHM of the
\oiii\ line:
\begin{equation}
{\rm FWHM}_{Avg} = (({\rm FWHM}_AF_A)^2+({\rm FWHM}_BF_B)^2)^\frac{1}{2}
\label{fwhma}
\end{equation}
where $F_A$ and $F_B$ are the fractional fluxes contained within the
two fitted components, A and B.  By primarily focussing on average
FWHMs, we avoid arbitrary definitions such as the threshold above
which a broad component constitutes an important contribution to the
overall flux of the emission line, or the point beyond which a
component is considered ``broad''.  This characterisation of the width
of the \oiii\ line also ensures that all AGNs, whether their \oiii\
lines fitted either with one or two gaussians, can be compared on an
equal footing (i.e., including, for example, AGNs in which the \oiii\
line is broad, but satisfactorily fit with a single gaussian).

By measuring the \oiii\ profiles of individual AGNs we can quantify
our sample in terms ${\rm FWHM_{Avg}}$.  In \fig{fractions} we show
the fractions of our sample with ${\rm FWHM_{Avg}}$ above a given
value.  It is clear that AGNs with relatively narrow \oiii\ lines
dominate our sample, with the two-thirds of all AGNs having ${\rm
  FWHM_{Avg}}<400$~\kms.  The median ${\rm FWHM_{Avg}}$ of
\loiii$>10^{42}$~\ergs\ AGNs is marginally higher than that of our
whole AGN sample ($\sim340$~\kms\ and $\sim330$~\kms), although this
may in part be due to a broad component being easier to identify among
the higher S/N spectra of these high \loiii\ AGNs.  Approximately $17$
per cent of all AGNs in our sample have ${\rm FWHM_{Avg}}>500$~\kms,
which is roughly the lower FWHM limit adopted by previous studies of
AGN outflows when selecting targets for follow-up resolved (i.e.,
long-slit or integral-field) spectroscopic studies of AGN outflows
(see Fig. 1 of \citealt{Harrison12}).  Also shown in \fig{fractions}
is the cumulative fraction for \lrad$>10^{23}$~\whz\ AGNs, which
indicates that the incidence of extremely broad \oiii\ lines (i.e.,
\fwhma$>1000$~\kms) is $\approx$5 times higher among these radio
luminous AGNs than among AGNs in general (although the fractions are
still small; e.g., $\sim5.5\%$ and $\sim1.1$, respectively, have ${\rm
  FWHM_{Avg}}>1000$~\kms).

In \fig{opt_dist_page} we plot the distribution of \fwhma\ as a
function of \loiii.  In the leftmost panel we plot all AGNs in our
sample (including both type 1 and type 2 AGNs).  As expected from the
cumulative fractions (\fig{fractions}), AGNs with
\fwhma$\lesssim500$~\kms\ dominate this plot at all \loiii\ values.
The upper envelope of this distribution increases with \loiii\ up to
\loiii$\sim10^{42}$~\ergs, although we reiterate that this may partly
be explained by the low S/N ratios of low luminosity AGNs' spectra
making it difficult to detect a broad emission line component.
Another possibility highlighted by the results of our spectral stacks
is that this increase is driven by the \loiii--\lrad\ correlation
coupled with an increase in \fwhma\ with \lrad.  To test for this, we
have split the sample according to \lrad\ in the other two panels of
\fig{opt_dist_page} (showing $<10^{23}$~\whz\ and $>10^{23}$~\whz in
the middle and right-hand panels, respectively).  We include AGNs with
upper limits on \lrad\ (provided they are $<10^{23}$~\whz) in the
middle panel, but do not include any upper limits in the right-hand
panel.  Compared to the whole sample (i.e., left-hand panel), there is
a distinct lack of \lrad$<10^{23}$~\whz\ AGNs with very broad \fwhma\
and the rise in upper envelope with \loiii\ is much less pronounced.
This difference is highly unlikely to be due to the smaller numbers in
the \lrad$<10^{23}$~\whz\ sample compared the whole sample, with a
KS-test indicating a $\ll0.1$ per cent chance that the two samples are
drawn at random from the same underlying population.

The right-hand panel of \fig{opt_dist_page} shows the \fwhma\ versus
\loiii\ distribution for \lrad$>10^{23}$~\whz.\footnote{We note that
  the numbers of AGNs in the middle and right-hand panels do not sum
  to the numbers in the left-hand panel as there are significant
  numbers of AGNs in our sample with \lrad\ upper limits
  $>10^{23}$~\whz\ which could not be reliably included in either the
  middle or right-hand plots.}  Despite there being far fewer sources
in this plot overall compared to the middle panel, over seven times
more of them lie at extreme values of \fwhma\ (i.e., $>1000$~\kms; 59
compared to 8) and the entire distribution is centred around a higher
\fwhma.  We find that \fwhma\ shows no clear dependence on \loiii\ for
these radio luminous AGNs.

\begin{figure*}
\begin{center}
	\includegraphics[width=15cm]{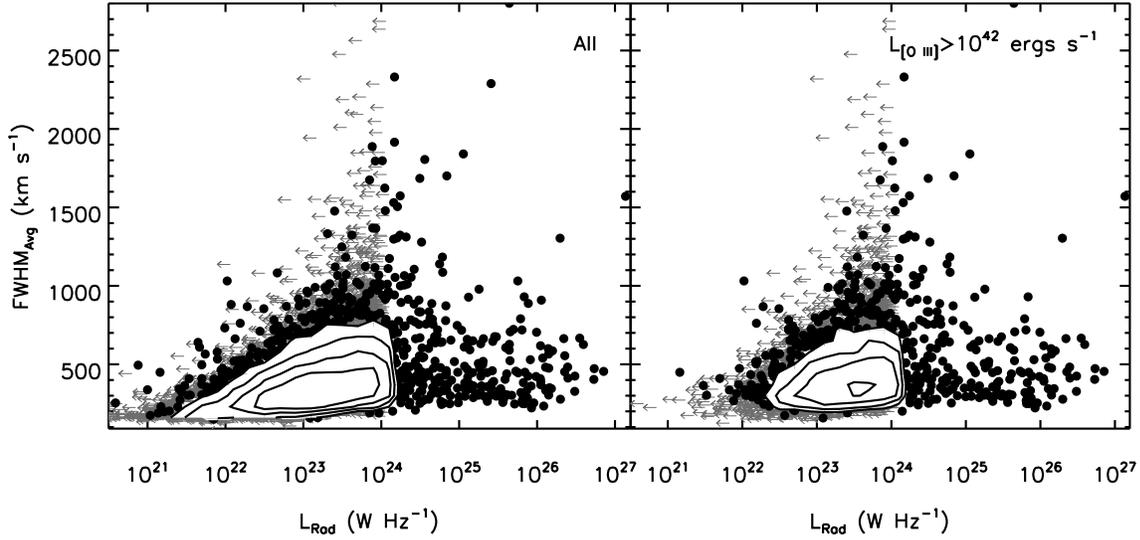}
\end{center}
\caption{Distribution of \fwhma\ plotted as a function of \lrad.  The
  left-hand panel includes all of the AGNs in our sample, whereas the
  right-hand panel show only those with \loiii$>10^{42}$~\ergs.  Both
  panels show a significant (i.e., ~0.05 per cent chance of being
  produced by random) peak in the upper envelope of the distribution
  around \lrad$\sim10^{24}$~\whz.}
\label{rad_dist_page}
\end{figure*}

In the left-hand panel of \fig{rad_dist_page} we plot \fwhma\ against
\lrad\ to further explore the connection between these two parameters.
We have included upper limits on \lrad\ which dominate the numbers at
\lrad$<10^{24}$~\whz.  The most striking aspect of this plot is the
peak in the \fwhma\ distribution around \lrad$=10^{23}-10^{25}$~\whz,
which are the \lrad\ associated with the broadest of our stacked
\oiii\ profiles.  As well as being traced by the detected sources,
this peak is also defined by the upper limits on the radio-undetected
sources.  Furthermore, a KS test indicates that this distribution
remains statistically the same if we restrict the sample to $z<0.123$,
at which point the flux limit of the corresponds to a maximum \lrad\
of $10^{23}$~\whz, indicating that the distribution is not produced by
a selection effect caused by the redshifts of the sources.  At first
sight this distribution suggests that moderate luminosity radio AGNs
are most affective at producing broad \oiii\ emission lines.  However,
other factors must first be ruled out before we can arrive at this
conclusion as it is possible that a combination of selection effects
could produce this distribution.

A possible explanation for the shape of the distribution at
\lrad$<10^{23}$~\whz\ is that the correlation between \lrad\ and
\loiii\ would make the broad component in these less radio luminous
AGNs more difficult to detect by our line fitting procedure due to the
weakness of the \loiii\ line.  However, when we reproduce the \fwhma\
versus \lrad\ plot using AGNs with \loiii$>10^{42}$~\ergs\ to ensure
that any broad component is detected (\fig{opt_dist_page} shows that
any such detection bias is negligible at these \oiii\ luminosities),
the same distribution is produced.  As such, we can rule out this
detection bias as having a major influence on the form of this
distribution.

The small numbers of AGNs in our sample with \lrad$>10^{25}$~\whz\ may
be the reason for the apparent drop-off in peak \fwhma\ at these high
\lrad.  However, a KS test reveals that this is unlikely to be the
case, returning only a $\sim$0.05 per cent probability that the
\fwhma\ of \lrad$=10^{24}-10^{25}$~\whz\ and $10^{25}-10^{26}$ AGNs
are drawn at random from the same underlying distribution.  This and
the arguments laid out above lead us to conclude that the peak in the
\fwhma--\lrad\ distribution around \lrad$=10^{23}-10^{25}$~\whz\ is
unlikely to be due to selection effects and thus gives a true
impression of the relationship between the profile of the \oiii\
emission line and the radio properties of the AGN.

\section{Discussion}
\label{Discussion}

In the previous sections we have explored how \oiii\ emission line
profiles differ between AGNs of different \lagn (from \loiii), \lrad,
\redd and \rrad.  Since emission line profiles are governed by the
kinematics of the emitting gas, our results allow us to assess how the
movement of gas surrounding the AGNs is related to these four key
observables.  In this section, we place our results in the context of
previous studies of the kinematics of the \oiii\ emitting gas and, in
doing so, consider the implications of our results on our
understanding of how AGN-driven outflows influence galaxy growth.

The results from both our spectral stacks and our individual \oiii\
line fits shows that, of the four observables considered (i.e., \lagn,
\lrad, \redd and \rrad), \lrad\ is the most important factor in
dictating the profile of this emission line.  Our results indicate
that the width of the \oiii\ line peaks between \lrad$=10^{23}$~\whz\
and $\sim10^{25}$~\whz\, with both more and less radio powerful AGNs
having, on average, narrower \oiii\ lines.  Similar results were
reported by \cite{Heckman84} and \cite{Whittle92} for much smaller
samples of AGNs.  The size and relative completeness of our sample
allows us to build on these previous studies, breaking the degeneracy
between \loiii\ and \lrad to identify the underlying links, and
placing AGNs with kinematically energetic narrow line regions in the
context of the general population by quantifying the fraction with
broad \oiii\ emission lines.

Spectral studies of small samples of AGNs displaying compact radio
jets have reported broadened \oiii\ emission lines that are co-spatial
with these jets (e.g., \citealt{Tadhunter03, Holt06, Holt08}),
suggesting that it is the jets' mechanical energy that is disturbing
the \oiii\ emitting gas.  Since we have no information regarding the
spatial extent of the \oiii\ emitting gas in our sample, we are unable
to perform such a direct analysis here.  Instead, we can exploit the
spatial resolution of the FIRST radio survey to determine whether AGNs
with broad \oiii\ have predominantly extended or compact radio
morphologies.  Taking NVSS-detected AGNs in our sample with
\lrad$=3\times10^{23}-3\times10^{24}$~\whz\ (to cover the peak of the
\fwhma\ distribution), we extract the angular extent of their radio
major axis from the cross-matched FIRST catalogue, which has a higher
spatial resolution at 1.4~GHz compared to NVSS.  We find all AGNs in
this sample with \fwhma$>1500$~\kms\ and single FIRST matches have
deconvolved extents $<2\arcsec$ (i.e., unresolved) and there is no
evidence that more extended radio sources have broader \oiii\ lines.
Furthermore, in the case of those with multiple FIRST matches, the
vast majority of AGNs in this subsample (i.e., $>80\%$) have at least
1 radio component closer than $1\arcsec$ to the galaxy nucleus,
suggestive of a radio core.  Finally, none of the AGNs in this
radio-selected subsample have multiple NVSS matches, meaning they are
unlikely to have extended, lobe-dominated radio emission without also
having a compact core.  While not conclusive, these various outcomes
support the view that compact (i.e., $\sim$kpc-scales or less) radio
cores play a major, if not the dominant, role in strongly disturbing
the \oiii\ emitting gas around optically-selected AGNs.  This may
  also explain why the \oiii\ width peaks at moderate \lrad, close to
  the division between Faranoff-Riley types I and II,
  since the latter dissipate more of their energy at larger
  (i.e., $\gg$~kpc) scales.  However, we also point out that the peak of
the \fwhma\ distribution is close to the point at which the radio
luminosity function becomes AGN, rather than star-formation,
dominated.  It is therefore feasible that the radio output of some of
the AGNs with the broadest \oiii\ lines is dominated by
star-formation.  

One factor that we have not yet explored is the possibility that some
of the AGN in our sample may have boosted radio emission due to
beaming, meaning that the intrinsic radio power may be considerable
lower than what we infer here.  However, while this may explain part
of the decreased disturbance toward higher \lrad, it is unlikely that
{\it all} (or even the majority) of the \lrad$\gtrsim10^{25}$~\whz\
AGNs in our sample are beamed (i.e., blazars), especially since
emission line selection would tend to bias {\it against} such AGNs as
they tend to be continuum dominated (e.g., \citealt{Morris91,
  Stocke91}).

Our analyses show a higher fraction of moderate radio-luminosity
(i.e., \lrad$=10^{23}-10^{25}$~\whz) AGNs display broad \oiii\ lines
compared to either weaker or stronger radio AGNs.  Furthermore, these
fractions are only weakly dependent on \lagn.  These moderate \lrad\
AGNs represent roughly 3 per cent of optically-selected AGNs at
$z<0.123$ (the redshift at which the NVSS survey is complete to
\lrad$>10^{23}$~\whz).  As such, while they constitute a small
fraction of optically-selected AGNs overall, they are at least
$\sim$10 times more common than the more radio luminous AGNs that are
typically the focus of detailed studies of AGN outflows (e.g.,
\citealt{Tadhunter03, Holt03, Nesvadba06, Holt08, Nesvadba08,
  Cano-Diaz12}; see Fig. 1 of \citealt{Harrison12} for a comparison).
Those studies that have sampled the \lrad$\sim10^{23}-10^{25}$~\whz,
\loiii$>10^{42}$~\ergs\ range have reported outflows extended over
galaxy (i.e., $\sim$few kpc) scales, suggesting that they could be
affecting their host galaxies (e.g., \citealt{Greene11,
  Harrison12}). A detailed comparison of the resolved radio components
and the gas kinematics is required to determine whether (and, if so,
how) the two are physically linked.

\section{Summary}
\label{Summary}
We have used a large (i.e., 24264) sample of optically selected AGNs
from the SDSS database to explore how the profiles of the forbidden
\oiii\ emission line relate to other key AGN parameters (bolometric
luminosity, Eddington ratio, radio loudness and radio luminosity) .
We use spectral stacking analyses to determine how the average \oiii\
profile changes as a function of these four parameters .  We also
explore the fractions of AGNs in each of these bins that have broad
\oiii\ emission lines.  Our main results can be summarised as:

\begin{itemize}
\item The average \oiii\ profile of Type 1 AGNs displays a prominent
  blue wing that can be well described by a broad (i.e.,
  FWHM$=851$~\kms) gaussian component that is blueshifted by 148~\kms\
  from a narrow (i.e., FWHM$1=335$~\kms) core.  The flux contained
  within this broad component represents 45 per cent of the total
  average line flux (see \S\ref{Average_Type1}).
\item When we separate the Type 1 AGNs into discrete bins of \loiii\
  or Eddington ratio (\redd) we find that the blue wings on the
  average \oiii\ profiles are increasingly prominent at higher values
  of \loiii\ and in our highest \redd bin (i.e., \redd$>0.3$;
  see \S\ref{Results:Bolometric} and \fig{stack_page}).  However, when
  we correct for the known correlation between \loiii\ and \lrad\ we
  find that \lrad\ has the most profound effect on the \oiii\ profile,
  with \lrad$=10^{23}-10^{25}$~\whz\ AGNs having the broadest
  profiles.   This appears to be the case for high Eddington ratio
    AGNs (i.e., \redd$>0.1$), although we are not able to show this
    conclusively in our highest \redd\ bin (\redd$>0.3$) due to the
    small number of such extreme AGN even in our large sample.
\item There is no difference in the average \oiii\ profiles of
  broad-line and narrow-line Seyfert 1 AGNs.
\item When the \oiii\ profiles of Type 2 AGNs are stacked in terms of
  \loiii\, their \oiii\ widths also increase with \loiii\, but are
  more symmetric than those of Type 1 AGNs.  We interpret this as
  suggesting that the blue wings are the result of outflowing material
  directed roughly along the axis of the obscuring torus.
\item Overall, roughly 25 per cent of all AGNs have average \oiii\
  FWHM (i.e., \fwhma) $>500$~\kms, compared to $\sim$50 per cent of
  AGNs with \lrad$>10^{23}$~\whz.  All of the AGNs in our sample with
  \fwhma$\gtrsim1200$~\kms\ have radio detections or upper limits
  consistent with \lrad$>10^{23}$~\whz.

\end{itemize}

Acknowledging the limiting spatial resolution of the FIRST and NVSS
surveys, our results are consistent with models in which compact radio
cores play an important role in disturbing the diffuse gas around the
AGN.  Ascertaining whether the impact of these cores is felt on
extended, galactic scales will require spatially resolved radio and
spectroscopic observations.

\section*{acknowledgements}
JRM and DMA acknowledges The Leverhulme Trust.  CMH acknowledges
funding from STFC.  This work has made use of data provided by the
SDSS.  Funding for the SDSS has been provided by the Alfred P. Sloan
Foundation, the Participating Institutions, the National Science
Foundation, the U.S. Department of Energy, the National Aeronautics
and Space Administration, the Japanese Monbukagakusho, the Max Planck
Society, and the Higher Education Funding Council for England. The
SDSS Web Site is http://www.sdss.org/.

\bsp

\appendix
\section{Multi-component fitting routine}
\label{linefit}

The emission line fitting procedure we have developed starts with a
simple, single gaussian fit to the $[$O~{\sc iii}$]\lambda\lambda
4959,5007$ emission lines and progresses by increasing the complexity
of the fit until each line is fit with multiple gaussians where
necessary.  At each stage of the fitting procedure any change in
\chisq\ is evaluated to establish whether the addition of any new
gaussian components has resulted in an improved fit (at a $>$99 per cent
confidence level).

First, the procedure attempts to fit each of the $[$O~{\sc
  iii}$]\lambda\lambda 4959,5007$ emission lines simultaneously with
single gaussians.  We require the FWHM and velocity offsets of these
gaussians to be equal throughout the fitting process and the
normalisation of the $[$O~{\sc iii}$]\lambda 4959$ gaussian fixed to
1/3 that of the \oiii\ gaussian (i.e., ${\rm FWHM_{[O~III]\lambda
    4959}} = {\rm FWHM_{[O~III]\lambda 5007}}$, $ v_{\rm
  [O~III]\lambda 4959} = v_{\rm [O~III]\lambda 5007}$, $ N_{\rm
  [O~III]\lambda 4959} = 1/3\cdot N_{\rm [O~III]\lambda 5007}$).  By
fitting $[$O~{\sc iii}$]\lambda4959$ we minimise any influence both
this and any neighbouring emission lines (e.g. the Fe~{\sc ii}
complexes) may have on the measurement of the \oiii\ profile.  We
note, however, that we do not use the results from fitting the
$[$O~{\sc iii}$]\lambda4959$ in our analysis.  The routine then
attempts to fit each of the permitted \hb\ and \ha\ lines
simultaneously with single gaussians (referred to later as the
`original' \hb\ and \ha\ gaussians), the velocity offsets and FWHM of
which are, again, linked (i.e., ${\rm FWHM_{H\beta}} = {\rm
  FWHM_{H\alpha}}$, $ v_{\rm H\beta} = v_{\rm H\alpha}$).  Second
gaussians are then added to the fits of the $[$O~{\sc
  iii}$]\lambda\lambda 4959,5007$ lines in an attempt to model any
complexities in the line profiles.  The velocity, FWHM and
normalisation of this second set of $[$O~{\sc iii}$]\lambda\lambda
4959,5007$ gaussians are linked in the same manner as the first set.
Finally, the procedure attempts to fit the narrow components of the
\hb\ and \ha\ lines and each of the $[$N~{\sc ii}$]\lambda\lambda
6548,6584$ lines with up to two further gaussians, depending on
whether the addition of the second $[$O~{\sc iii}$]\lambda\lambda
4959,5007$ gaussians resulted in a significantly improved fit.  Each
of these additional gaussians are forced to have the same velocity
offsets and FWHM as their respective $[$O~{\sc iii}$]\lambda\lambda
4959,5007$ counterparts, but their relative normalisations are allowed
to vary.  We caution, however, that it has previously been reported
that the line profiles of different forbidden species can differ
(e.g., \citealt{Komossa08}).  By adopting the approach that the
relative normalisation of the two \oiii-derived narrow-line components
can vary, we adopt a compromise between having some reasonable
constraints on the fit to the permitted and \nii\ lines (thus
preventing these lines from broadening to accommodate the true broad
permitted lines), while allowing the profile to vary somewhat to
accommodate some of the differences between different species.  The
normalisation of the $[$N~{\sc ii}$]\lambda\lambda 6548$ line is fixed
to 1/3 that of the \nii\ line.  In fitting the narrow components of
the \hb\ and \ha\ lines with the modelled $[$O~{\sc
  iii}$]\lambda\lambda 4959,5007$ profile, the original \hb\ and \ha\
gaussians are allowed to widen to fit any broad components of these
permitted lines.

\begin{figure*}
\begin{center}
	\includegraphics[width=16cm]{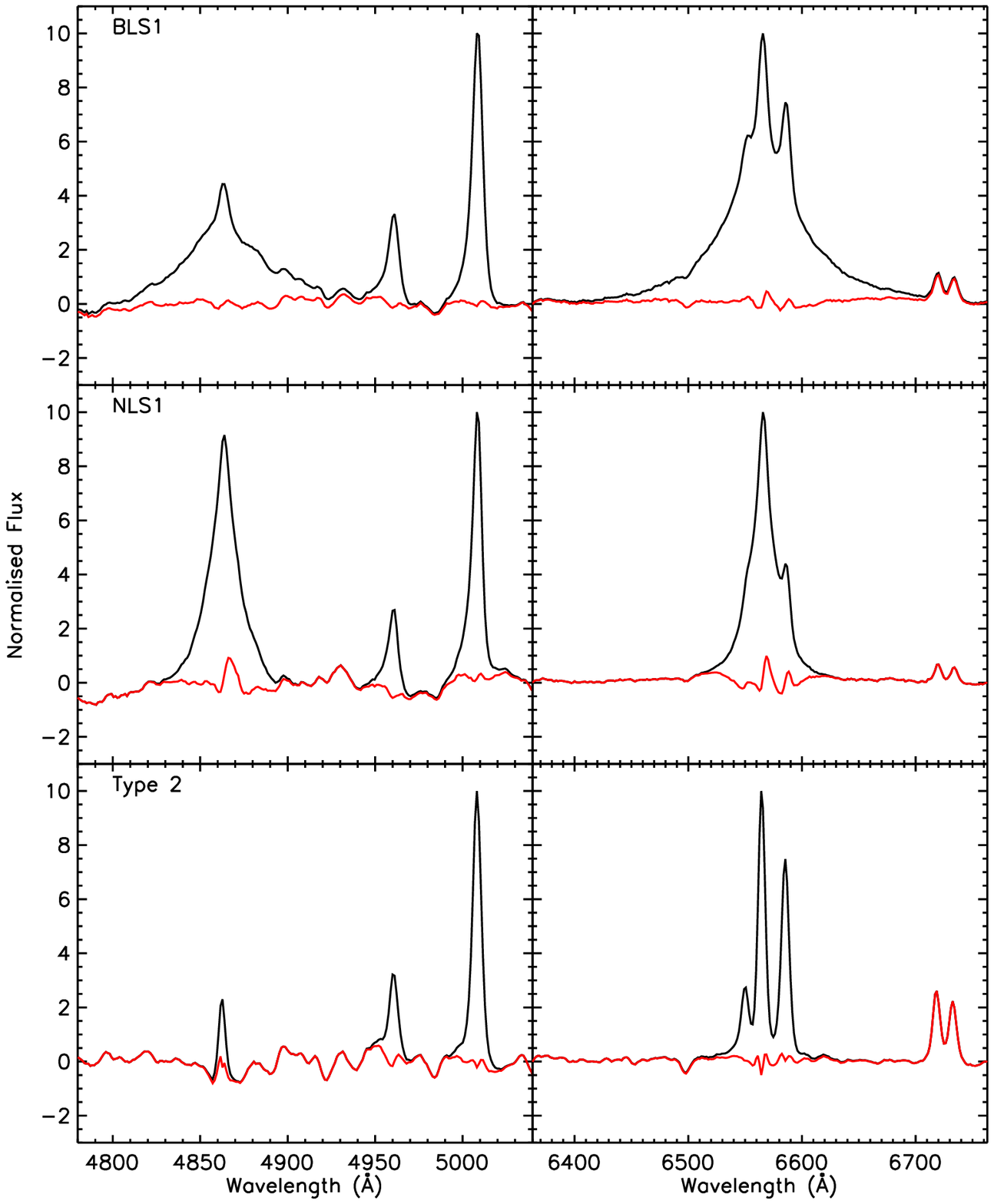}
\end{center}
\caption{Stacked emission line profiles ({\it left}: \hb, \oiii; {\it
    right}: \ha) for 1000 BLS1s ({\it top}), NLS1s ({\it middle}) and
  Type 2s ({\it bottom}) AGNs randomly selected from our full sample.
  The red lines show the results of stacking the spectra after
  subtracting the best fit line model (i.e.,
  $\sum\limits_{i}(F_i(\lambda) - F_i^{\rm model}(\lambda))$.  We take
  the general lack of strong systematics in these average residual
  spectra as evidence that, over the sample, our line fitting routine
  is doing a reasonable job of modelling the profiles of the emission
  lines, particularly the broad wings of the \ha\ and \hb\ lines and
  the overall \oiii\ profile.}
\label{spec_less_model}
\end{figure*}

Since the goal of our fitting the emission line profiles is to look
for general trends within the large AGN population, we are not greatly
concerned if a small number of ``one off'' spectra are poorly fit by
our routine, provided that it provides accurate fit parameters for the
vast majority of sources in our sample. To demonstrate that this is
the case, we plot the results of stacking randomly selected spectra
after subtracting the best fit model (see the red line in
\fig{spec_less_model}). We perform this for separate samples of BLS1s,
NLS1s and Type 2 AGNs, randomly selecting 1000 AGN from each of these
three samples. In general, we see only weak residuals around the
fitted emission lines, especially the \oiii\ lines (the bump around
the \oiii\ line in the NLS1 average spectrum is due to the Fe~{\sc ii}
complex which is stronger, in general, in this class AGNs). There is
evidence of a narrow residual in the \ha\ and \hb\ lines of the NLS1s
(with a peak roughly 1/10th the maximum flux of these lines). However,
since we only use the broad components of these lines to discriminate
between NLS1s and BLS1s, which are generally well fit, this will not
affect our main results. As such, we take these average residuals as
evidence that, in general, our line fitting routine is providing an
accurate and unbiased parameterisation of the emission profiles of our
sample spectra.

\label{lastpage}

\end{document}

%% file: Table1.tex
\begin{tabular}{@{}ccccc@{}}
\hline
\hline
$L_{\rm Rad}$&$L_{\rm [O III]}$&$N$&Med. $L_{\rm Rad}$&Med. $L_{\rm [O III]}$\\
(1)&(2)&(3)&(4)&(5)\\
\hline
$<10^{23}$&$10^{40}-10^{41}$&    129&$<23$&40.8\\
$<10^{23}$&$10^{41}-10^{42}$&    129&$<23$&41.6\\
$<10^{23}$&$10^{42}-10^{43}$&    129&$<23$&42.3\\
&&&&\\
$>10^{23}$&$10^{41}-10^{42}$&     64&24.1&41.7\\
$>10^{23}$&$10^{42}-10^{43}$&     77&24.0&42.5\\
$>10^{23}$&$10^{43}-10^{44}$&     77&24.1&43.3\\
&&&&\\
$<10^{22}$&$<10^{42}$&     33&$<22$&41.5\\
$10^{22}-10^{23}$&$<10^{42}$&     33&22.6&41.7\\
$10^{23}-10^{24}$&$<10^{42}$&     32&23.4&41.8\\
$10^{24}-10^{25}$&$<10^{42}$&     33&24.3&41.7\\
&&&&\\
$<10^{22}$&$>10^{42}$&     39&$<22$&42.1\\
$10^{22}-10^{23}$&$>10^{42}$&     42&22.7&42.3\\
$10^{23}-10^{24}$&$>10^{42}$&     42&23.5&42.4\\
$10^{24}-10^{25}$&$>10^{42}$&     42&24.3&42.4\\
$10^{25}-10^{26}$&$>10^{42}$&     37&25.5&42.4\\
\hline
\end{tabular}